\documentclass[useAMS,usenatbib]{mn2e}
\usepackage{graphicx}
\usepackage{amsmath}

\title[Particle dynamics in
self-gravitating discs]{Planetesimal formation in self-gravitating
discs -- the effects of particle self-gravity and back-reaction}

\author[Gibbons et al.]{P. G. Gibbons$^{1}$\thanks{E-mail: pgg@roe.ac.uk}, G. R. Mamatsashvili$^{2}$ and W. K. M. Rice$^{1}$\\
$^{1}$SUPA, Institute for Astronomy, Royal Observatory, Blackford Hill, Edinburgh EH9 3HJ\\
$^{2}$Department of Physics, Faculty of Exact and Natural Sciences,
Tbilisi State University, Il. Chavchavadze ave. 3, Tbilisi 0179,
Georgia}

\begin{document}

\date{Accepted ?. Received ? ; in original form ?}

\pagerange{\pageref{firstpage}--\pageref{lastpage}} \pubyear{2013}

\maketitle

\label{firstpage}

\begin{abstract}
We study particle dynamics in self-gravitating gaseous discs with a
simple cooling law prescription via two-dimensional simulations in
the shearing sheet approximation. It is well known that structures
arising in the gaseous component of the disc due to a gravitational
instability can have a significant effect on the evolution of dust
particles. Previous results have shown that spiral density waves can
be highly efficient at collecting dust particles, creating
significant local over-densities of particles. The degree of such
concentrations has been shown to be dependent on two parameters: the
size of the dust particles and the rate of gas cooling. We expand on
these findings, including the self-gravity of dust particles, to see
how these particle over-densities evolve. We use the
{\scriptsize{PENCIL CODE}} to solve the local shearing sheet
equations for gas on a fixed grid together with the equations of
motion for solids coupled to the gas through an aerodynamic drag
force. We find that the enhancements in the surface density of
particles in spiral density wave crests can reach levels high enough
to allow the solid component of the disc to collapse under its own
self-gravity. This produces many gravitationally bound collections
of particles within the spiral structure. The total mass contained
in bound structures appears nearly independent of the cooling time,
suggesting that the formation of planetesimals through dust particle
trapping by self-gravitating density waves may be possible at a
larger range of radii within a disc than previously thought. So,
density waves due to gravitational instabilities in the early stages
of star formation may provide excellent sites for the rapid
formation of many large, planetesimal-sized objects.

\end{abstract}


\section{Introduction}

The field of planet formation currently provides two methods through
which large gas giant planets can form in discs around young stars.
The favoured model of planet formation is known as the core
accretion model. This model proposes that planets grow via a
`bottom-up' process, where a core of solid material grows from
initially small, kilometre-sized objects via a series of collisions.
If this core becomes massive enough, it will begin to accrete a
gaseous envelope from the disc \citep{Pollack1996}. For a
Jupiter-like gas giant planet to form,  the solid core must reach a
mass of  $\sim 10$ Earth masses before the disc is depleted of gas,
a process which is observationally estimated to take from $10^6$ to
$10^7$ years \citep{Haisch2001}. A key uncertainty in the core
accretion model is the mechanism through which the disc becomes
populated with kilometre-sized solid objects, similar to those found
in the asteroid belt. It is likely that these objects are assembled
via collisional growth from initial small dust grains present in the
Interstellar Medium (ISM) during the star formation process that
creates the protoplanetary disc. However, current theory has
difficulties explaining the growth of dust grains past the
metre-scale -- the velocities of metre-sized objects should be
larger than the critical threshold for sticking \citep[see
e.g.,][]{BlumWurm2008,Guttler2010}, so individual collisions between
the bodies are no longer expected to be constructive. In this case,
the self-gravity of any resulting rubble pile will be too weak to
allow the debris to collapse into a gravitationally bound structure.

The dynamics of these smaller particles that ultimately grow to form
kilometre-sized planetesimals is largely governed by the aerodynamic
drag force that arises from the velocity difference between the
particles and the surrounding gas. The radial pressure gradient
within the disc tends to be negative, making the gas orbit with
sub-Keplerian velocities. The dust is not affected by the gas
pressure gradient and would orbit at Keplerian velocities in the
absence of drag. The drag force exerted on the dust results in the
solids losing angular momentum to the disc and drifting inward at a
rate that depends on the particles' size \citep{Weid1977}. For very
small grain sizes, the dust is tightly coupled to the gas in the
disc and the radial drift velocities are small. For very large
objects, the solids are decoupled from the gas, move in
approximately Keplerian orbits and again have very small drift
velocities. Particles in the intermediate size range can, however,
have large drift velocities. Although the exact size range depends
on the local properties of the disc, drift velocities can exceed
$10^3$cm/s for objects with sizes between 1 cm and 1 m
\citep{Weid1977}. Therefore the process through which planetesimals
form must be rapid, unless this inward drift is offset.
\citet{Laibe2012} have shown that there may be surface density and
temperature profiles for which particles may survive this inward
migration, and \citet{Rice2004,Rice2006,Gibbons2012} have shown that
local pressure maxima associated with density waves due to
gravitational instabilities in the disc can trap the particles,
saving them from the inward drift. Nevertheless, in the standard
core accretion scenario, the period of growth from micron- to
decametre-sized objects is assumed to be rapid, otherwise objects in
this size range would rapidly spiral inward and be accreted onto the
central protostar.

In thin discs, gravitational instabilities are characterized by the
\cite{Toomre1964} parameter,
\[
Q = \frac{c_s\Omega}{\pi G\Sigma},
\]
where $c_s$ is the gas sound speed, $\Omega$ is the Keplerian
rotation frequency and $\Sigma$ is the disc surface density.
Axisymmetric instability occurs for $Q<1$, while non-axisymmetric
one can emerge for $Q<1.5-1.7$ \citep{Durisen2007}. If a disc is
susceptible to such instabilities, depending on the thermal
properties of the disc, one of two outcomes may occur. If the
cooling time is greater than some threshold, $t_{c, crit}$, the disc
will settle into a quasi-steady state, where the cooling balances
the heating generated by gravitoturbulence \citep{Gammie2001}. For
cooling times shorter than $t_{c, crit}$, the disc may fragment,
forming brown dwarf and/or gas giant planet type objects
\citep{Boss1998}. The critical cooling time below which
fragmentation occurs is commonly taken to be $t_{c,crit}
=3\Omega^{-1}$ \citep{Gammie2001, Rice2003}, however recent studies
suggest that this threshold may not be fully converged with recent
high resolution simulations, indicating that the critical cooling
time, $t_{c,crit}$, may even exceed $10\Omega^{-1}$
\citep{Meru2011}. It has, however, been suggested
\cite{Paardekooper2011, Lodato2011, Rice2012} that this
non-convergence is a numerical issue rather than actually suggesting
that fragmentation could typically occur for $t_c > 10 \Omega^{-1}$.
\citet{Paardekooper2012} do, however, suggest that there may be an
intermediate range of cooling times for which fragmentation may
indeed be stochastic, observing fragmentation in some simulations
with cooling times as high as $t_c = 20\Omega^{-1}$. Although very
few Class II objects are observed to have sufficiently massive discs
for gravitational instabilities to set in \citep{Beckwith1991},
observations indicate that during the Class 0 and Class I phases,
massive discs are much more common \citep{Rodriguez2005,
Eisner2005}, suggesting that most, if not all stars possess a
self-gravitating disc for some period of time during the earliest
stages of star formation. If this is the case, these instabilities
will likely take the form of non-axisymmetric spiral structures.

It has been shown that these spiral waves are highly effective at
trapping the solids in the disc. \citet{Rice2004} showed, using
global disc simulations, that the surface density of certain
particle sizes can be enhanced by a factor of over 100 in spiral
wave structure. \citet[][Paper I]{Gibbons2012} used local
shearing-sheet simulations to expand on these findings, mimicking
the conditions at a range of disc radii to study the particle
trapping capabilities of spiral density waves through the disc.
These results showed that gravitational instabilities are
responsible for creating large over-densities in the solid component
of the disc at intermediate to large orbital radii $(>20\rm{AU})$
within the disc. \citet{Rice2006} estimated from global disc
simulations that the observed increase in the surface density of
solids will lead to the creation of kilometre-scale planetesimals
due to the gravitational collapse of the solids in these over-dense
regions. Here we aim to extend this to study the gravitational
collapse of the solids via local shearing-sheet simulations.

The goal of the present work is to demonstrate how the
over-densities that form in the solid component of the disc can
undergo gravitational collapse as a result of the solids'
self-gravity, promoting further grain growth, which can ultimately
lead to the formation of planetesimals at very early evolutionary
stages when the disc is still self-gravitating. This directly
expands on the work in Paper I, where we studied the effect of
varying effective cooling time of the gas on the particle-trapping
capabilities of spiral density waves for a range of particle sizes
(friction times), but including neither the particle self-gravity
nor the back-reaction from the particles on gas via drag force. In
this paper, taking into account both these factors, we numerically
studied dynamical behaviour of particles embedded in a
self-gravitating disc using a local shearing sheet approximation. We
investigated the possibility that density enhancements in the solid
component of the disc can lead to the direct formation of
gravitationally bound solid clumps and, if so, study how such clumps
might behave. In particular, we are interested in whether
gravitationally bound accumulations of solids can form within the
disc, since the formation of a large reservoir of planetesimals at
early times in the disc is a major obstacle for the core accretion
theory. In this regard, we would like to mention that self-gravity
of the solid component has been demonstrated to be a principal agent
promoting the formation of large planetesimals inside gaseous
over-densities arising in compressible magnetohydrodynamic
turbulence driven by the magnetorotational (MRI) instability in
discs \citep{Johansen2007, Johansen2011}.

The paper is organized as follows. In Section 2 we outline the disc
model and equations we solve in our simulations. In Section 3 we
describe the evolution of the gas and dust particles. Summary and
discussions are given in Section 4.
\\

\section[]{Dynamical equations}
\label{Model}

To investigate the dynamics of solid particles embedded in a
self-gravitating protoplanetary disc, we solve the two-dimensional
(2D) local shearing sheet equations for gas on a fixed grid,
including disc self-gravity as in \citet{Gammie2001}, together with
the equations of motion of solid particles coupled to the gas
through aerodynamic drag force. As mentioned in the Introduction,
following \cite{Johansen2011}, we also include self-gravity of
particles to examine their collapse properties. As a main numerical
tool, we employ the {\scriptsize{PENCIL CODE}} \footnote{See
http://code.google.com/p/pencil-code/}. The {\scriptsize{PENCIL
CODE}} is a sixth order spatial and third order temporal finite
difference code (see \citet{Brandenburg2003} for full details). The
{\scriptsize{PENCIL CODE}} treats solids as numerical
super-particles \citep{Johansen2006,Johansen2011}.

In the shearing sheet approximation, disc dynamics is studied in the
local Cartesian coordinate frame centred at some arbitrary radius,
$r_0$, from the central object and rotating with the disc's angular
frequency, $\Omega$, at this radius. In this frame, the $x$-axis
points radially away from the central object, the $y$-axis points in
the azimuthal direction of the disc's differential rotation, which
in turn manifests itself as an azimuthal parallel flow characterized
by a linear shear, $q$, of background velocity along the $x-$axis,
${\bf u}_0=(0,-q\Omega x)$. The equilibrium surface densities of
gas, $\Sigma_0$, and particles, $\Sigma_{p,0}$, are spatially
uniform. Since the disc is cool and therefore thin, the aspect ratio
is small, $H/r_0\ll 1$, where $H=c_s/\Omega$ is the disc scale
height and $c_s$ is the gas sound speed. The local shearing sheet
model is based on the expansion of the basic 2D hydrodynamic
equations of motion to the lowest order in this small parameter
assuming that the disc is also razor thin \citep[see
e.g.,][]{Gammie2001}.

Our simulation domain spans the region $-L_x/2 \leq x \leq L_x/2$,
$-L_y/2 \leq y \leq L_y/2$. As is customary, we adopt the standard
shearing-sheet boundary conditions \citep{Hawley1995}, namely for
any variable $f$, including azimuthal velocity with background flow
subtracted, we have
\[
f(x,y,t) = f(x+L_x,y-q\Omega L_xt,t),~~~~~~(x-{\rm boundary})
\]
\[
f(x,y,t) = f(x,y+L_y,t),~~~~~~~(y-{\rm boundary})
\]
The shear parameter $q=1.5$ for the Keplerian rotation profile
adopted in this paper.

\subsection{Gas density}

In this local model, the continuity equation for the vertically
integrated gas density $\Sigma$ is
\begin{equation}
\frac{\partial\Sigma}{\partial t} + {\bf\nabla}\cdot(\Sigma{\bf u})
-q\Omega x\frac{\partial\Sigma}{\partial y}-f_D(\Sigma) = 0
\end{equation}
where ${\bf u}=(u_x,u_y)$ is the gas velocity relative to the
background Keplerian shear flow ${\bf u}_0$. Due to the high-order
numerical scheme of the {\scriptsize{PENCIL CODE}} it also includes
a diffusion term, $f_D$, to ensure numerical stability and capture
shocks,
\[
f_D = \zeta_D(\nabla^2 \Sigma +\nabla \textrm{ ln } \zeta_D \cdot
\nabla\Sigma).
\]
Here the quantity $\zeta_D$ is the shock diffusion coefficient
defined as
\[
\zeta_D = D_{sh} \langle \max_3[(-\nabla\cdot {\bf u})_+]
\rangle(\Delta x)^2 \label{shock}
\]
where $D_{sh}$ is a constant defining the strength of shock
diffusion as outlined in Appendix B of \citet{Lyra2008a}. $\Delta x$
is the grid cell size.

\subsection{Gas velocity}

The equation of motion for the gas relative to the unperturbed
Keplerian flow takes the form
\begin{multline}
\frac{\partial {\bf u}}{\partial t}+({\bf u}\cdot\nabla){\bf
u}-q\Omega x \frac{\partial {\bf u}}{\partial y} = -\frac{\nabla
P}{\Sigma} - 2\Omega {\bf\hat{z}}\times{\bf u}+q\Omega u_x
{\bf\hat{y}}\\-\frac{\Sigma_p}{\Sigma}\cdot\frac{{\bf u}-{\bf
v}_p}{\tau_f}+2\Omega\Delta v{\bf \hat{x}}- \nabla\psi +\bf f_\nu
(u), \label{gvel1}
\end{multline}
where $P$ is the vertically integrated pressure, $\psi$ is the
gravitational potential produced together by the perturbed gas
surface density, $\Sigma-\Sigma_0$, and the vertically integrated
bulk density of particles, $\Sigma_p-\Sigma_{p,0}$ (see equation 6
below). The left hand side of equation (2) describes the advection
by the velocity field, {\bf u}, itself and by the mean Keplerian
flow. The first term on the right hand side is the pressure force.
The second and third terms represent the Coriolis force and the
effect of shear, respectively. The fourth term describes the
aerodynamic drag force, or back-reaction exerted on the gas by the
dust particles \citep[see e.g.,][]{Lyra2008b,Lyra2009,Johansen2011}.
This force depends on the difference between the velocity of
particles ${\bf v}_p$ and the gas velocity and is inversely
proportional to the stopping, or friction time, $\tau_f$, of
particles. The fifth term mimics a global radial pressure gradient
which reduces the orbital speed of the gas by the positive amount
$\Delta v$ and is responsible for the inward radial migration of
solids in an unperturbed disc. The sixth term represents the force
due to self-gravity of the system. Finally, the code includes an
explicit viscosity term, $\bf f_\nu$,
\begin{align*}
{\bf f_\nu} =&  \nu(\nabla^2{\bf u} + \frac{1}{3}\nabla\nabla\cdot{\bf u}
+ 2{\bf S}\cdot \nabla \textrm{ln }\Sigma) \nonumber \\
& + \zeta_\nu[\nabla(\nabla\cdot{\bf u})+ (\nabla \textrm{ln }\Sigma
+ \nabla\textrm{ln }\zeta_\nu)\nabla\cdot{\bf u}],
\end{align*}
which contains both Navier-Stokes viscosity and a bulk viscosity for
resolving shocks. Here {\bf S} is the traceless rate-of-strain
tensor
\[
S_{i j} = \frac{1}{2}\left(\frac{\partial u_i}{\partial
x_j}+\frac{\partial u_j}{\partial x_i} - \frac{2}{3}\delta_{i
j}\nabla\cdot{\bf u}\right)
\]
and $\zeta_{\nu}$ is the shock viscosity coefficient analogous to
the shock diffusion coefficient $\zeta_D$ defined above, but with
$D_{sh}$ replaced by $\nu_{sh}$.

\subsection{Entropy}

The {\scriptsize{PENCIL CODE}} uses entropy, $s$, as its main
thermodynamic variable, rather than internal energy, $U$, as used by
\citet{Gammie2001}. The equation for entropy evolution is
\begin{equation}
\frac{\partial s}{\partial t}+({\bf u}\cdot\nabla)s - q\Omega
x\frac{\partial s}{\partial y} = \frac{1}{\Sigma
T}\left(2\Sigma\nu{\bf S}^2 - \frac{\Sigma
c_s^2}{\gamma(\gamma-1)t_c} + f_{\chi}(s)\right)
\end{equation}
where the first term on the right hand side is the viscous heating
term and the second term is an explicit cooling term. Here we assume
the cooling time $t_c$ to be constant throughout the simulation
domain and take its value to be sufficiently large that the disc
does not fragment and achieves a quasi-steady state. The final term
on the right hand side,  $f_{\chi}(s)$, is a shock dissipation term
analogous to that outlined for the density.

\subsection{Dust particles}

The dust particles are treated as a large number of numerical
super-particles \citep{Johansen2006,Johansen2011} with positions
${\bf x}_p=(x_p,y_p)$ on the grid and velocities ${\bf v}_p=({\rm
v}_{p,x},{\rm v}_{p,y})$ relative to the unperturbed Keplerian
rotation velocity, ${\bf v}_{p,0}=(0,-q\Omega x_p)$, of particles in
the local Cartesian frame. These are evolved according to
\begin{equation}
\frac{\mathrm{d}{\bf x}_p}{\mathrm{d}t} = {\bf v}_p - q\Omega
x_p{\bf \hat{y}}
\end{equation}
\begin{equation}
\frac{\mathrm{d}{{\bf v}_p}}{\mathrm{d}t} = - 2\Omega
{\bf\hat{z}}\times{\bf v}_p+q\Omega {\rm v}_{p,x}
{\bf\hat{y}}-\nabla\psi+\frac{{\bf u} - {\bf v}_p}{\tau_f}.
\label{parvel}
\end{equation}
The first two terms on the right hand side of equation (5) represent
the Coriolis force and the non-inertial force due to shear. The
third term is the force exerted on the particles due to the common
gravitational potential $\psi$. The fourth term describes the drag
force exerted by the gas on the particles which arises from the
velocity difference between the two. Unlike the gas, the particles
do not feel the pressure force. In the code, the drag force on the
particles from the gas is calculated by interpolating the gas
velocity field to the position of the particle, using the second
order spline interpolation outlined in Appendix A of
\citet{Youdin2007}. The back-reaction on the gas from particles in
equation (2) is calculated by the scheme outlined in
\citet{Johansen2011}.

\subsection{Self-gravity}
The gravitational potential in the dynamical equations (2) and (5)
is calculated by inverting Poisson equation for it, which contains
on the right hand side the gas plus particle surface densities in a
razor thin disc \citep[e.g.,][]{Lyra2009}
\begin{equation}
\Delta\psi = 4\pi
G(\Sigma-\Sigma_0+\Sigma_p-\Sigma_{p,0})\delta(z)\label{Poisson}
\end{equation}
using the Fast Fourier Transform (FFT) method outlined in the
supplementary material of \citet{Johansen2007}. Note that the
perturbed gas, $\Sigma-\Sigma_0$, and particle,
$\Sigma_p-\Sigma_{p,0}$, surface densities enter equation (6), since
only the gravitational potential associated with the perturbed
motion (and hence density perturbation) of both the gaseous and
solid components determine gravity force in equations (2) and (5).
Here, the surface density is Fourier transformed from the
$(x,y)$-plane to the $(k_x,k_y)$-plane without the intermediate
co-ordinate transformation performed by \citet{Gammie2001}. For this
purpose, a standard FFT method has been adapted to allow for the
fact that the radial wavenumber $k_x$ of each spatial Fourier
harmonic depends on time as $k_x(t) = k_x(0) + q\Omega k_yt$ in
order to satisfy the shearing sheet boundary conditions \citep[see
also][]{Mamatsashvili2009}.

\subsection{Units and initial conditions}

We normalise our parameters by setting $c_{s0}=\Omega=\Sigma_0= 1$.
The time and velocity units are $[t] = \Omega^{-1}$ and $[u] =
c_{s0}$, resulting in the orbital period $T = 2\pi$. The unit of
length is the scale-height, $[l] = H = c_{s0}/\Omega$. The initial
Toomre $Q=c_{s0}\Omega/\pi G\Sigma_0$ parameter is taken to be 1
throughout the domain. This sets the gravitational constant $G$ =
$\pi^{-1}$. The surface density of gas is initially uniform and set
to unity. The simulation domain is a square with dimensions
$L_x=L_y=80G\Sigma_0/\Omega^2$ and is divided into a grid of
$N_x\times N_y=1024\times 1024$ cells with sizes $\Delta x\times
\Delta y=L_x/N_x\times L_y/N_y$. This choice of units sets the
domain size $L_x = 80H/\pi Q=25.46H$. It is worth noting that the
cooling time, $t_c$, which we have assumed to be constant throughout
the sheet, in reality is $t_c = t_c(\Sigma,U,\Omega)$ as described
by \citet{Johnson2003}. However, the use of constant cooling time
over a sheet of this size allows us to infer the general behaviour
of the dust particles at a given location within the disc.

The gas velocity field is initially perturbed by some small random
fluctuations with the uniform rms amplitude $ \sqrt{\langle\delta
{\bf u}^2\rangle} = 10^{-3}$. We take the viscosity and diffusion
coefficients to be $\nu = 10^{-2}$ and $\nu_{sh} = D_{sh} = 5.0$. As
shown in Paper I, typical values of the radial drift parameter,
$\Delta v$, does not have a significant effect on the outcome of the
simulations, therefore in all the simulations presented below we
take $\Delta v=0.02$. We use $5\times10^5$ particles, split evenly
between five friction times, $\tau_f =
[0.01,0.1,1,10,100]\Omega^{-1}$. \citet{Bai2010} and
\citet{Laibe2012} have shown that there is a spatial resolution
criteria which applies to coupled dust and gas simulations such as
those outlined above. For the dust particles to be properly
resolved, the grid spacing must satisfy $\Delta x < c_s\tau_f$. For
the chosen set of parameters we have $\Delta x \sim 0.07c_s\tau_f$,
so this condition is satisfied for all but the $\tau_f =
0.01\Omega^{-1}$ particles. As noted in Paper I, this
under-resolution of particles does not appear to create any
numerical inconsistencies in the evolution of the smallest
particles.

In all the runs below, each particle species with a fixed
radius/friction time is distributed spatially uniformly with the
average surface density of $\Sigma_{p,0}=10^{-2}\Sigma_0$ prescribed
according to the standard value of dust-to-gas ratio, except one low
particle mass run (Fig. 1), where we take
$\Sigma_{p,0}=10^{-3}\Sigma_0$. The particles are initially given
random positions within the sheet and zero velocities, relative to
the background Keplerian flow, ${\bf v}_p(t=0)=0$.

\section{Results}

\subsection{Gas evolution}

The evolution of the gaseous component of the disc is in good
agreement with that observed in analogous studies based on the
shearing sheet formalism \citep[][Paper
I]{Gammie2001,Johnson2003,Mamatsashvili2009, Rice2011}. The small
initial velocity fluctuations grow and develop into non-linear
fluctuations in velocity, surface density and potential. Shocks then
develop which proceed to heat the gas, while the cooling acts to
reduce the entropy of the gas. Density structures develop which are
sheared out by differential rotation. These density structures tend
to take a trailing form, leading to a finite shear stress and
angular momentum transport parameter $\alpha$ \citep{Gammie2001}.
After a few orbits, the heating due to shocks is balanced by the
cooling term and the disc settles into a quasi-steady self-regulated
gravitoturbulent state (Fig. 1, left panel), where the
domain-averaged thermal, kinetic and gravitational energies of the
disc are approximately constant with time. In this state, the
surface density field clearly shows elongated trailing surface
density features, or spiral density waves. The amplitude of these
density waves, and consequently their capability to trap dust
particles, depends on the cooling time of the gas
\citep{Cossins2009,Rice2011}.

\begin{figure*}
\includegraphics[width = 0.49\textwidth]{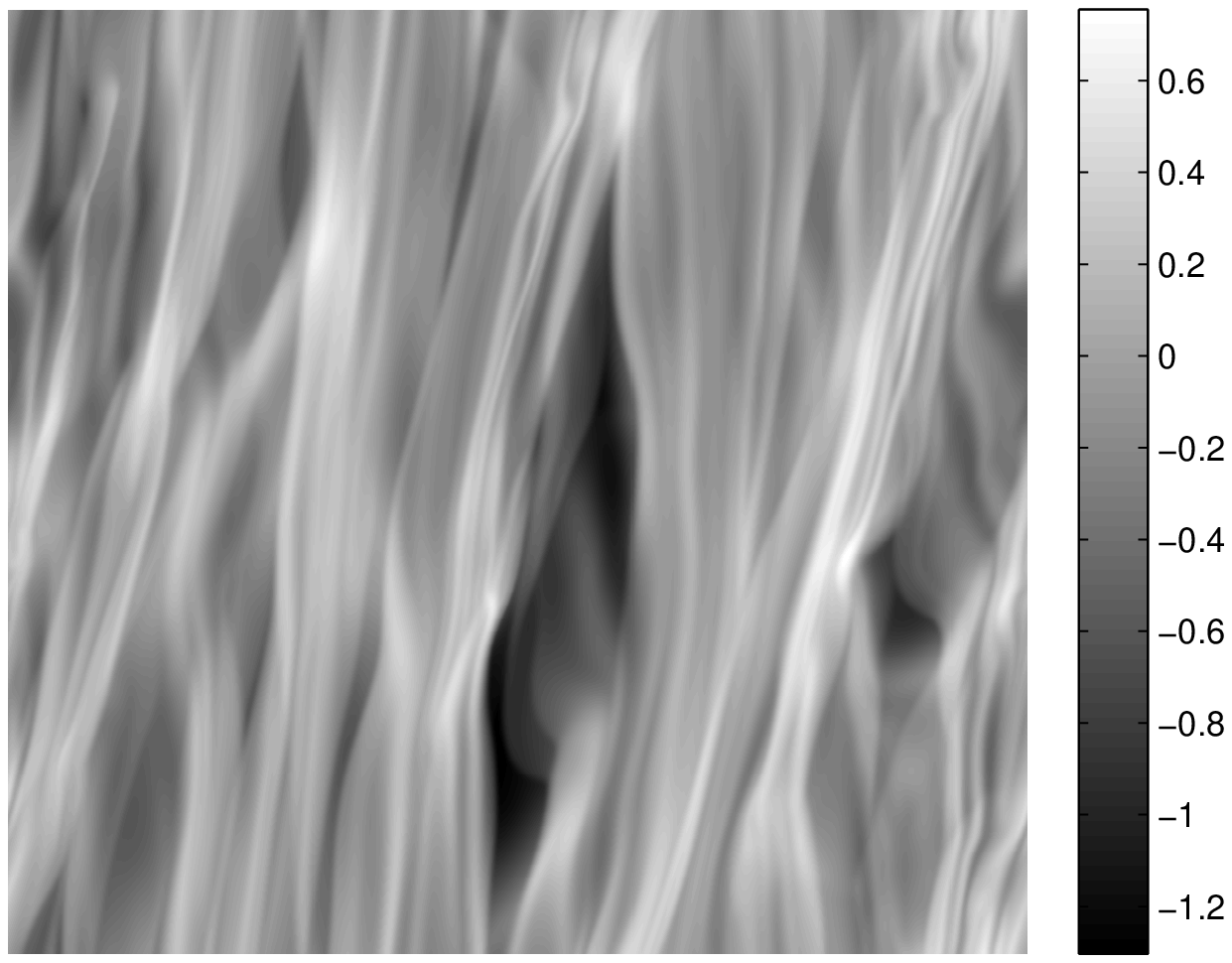}
\includegraphics[width = 0.49\textwidth]{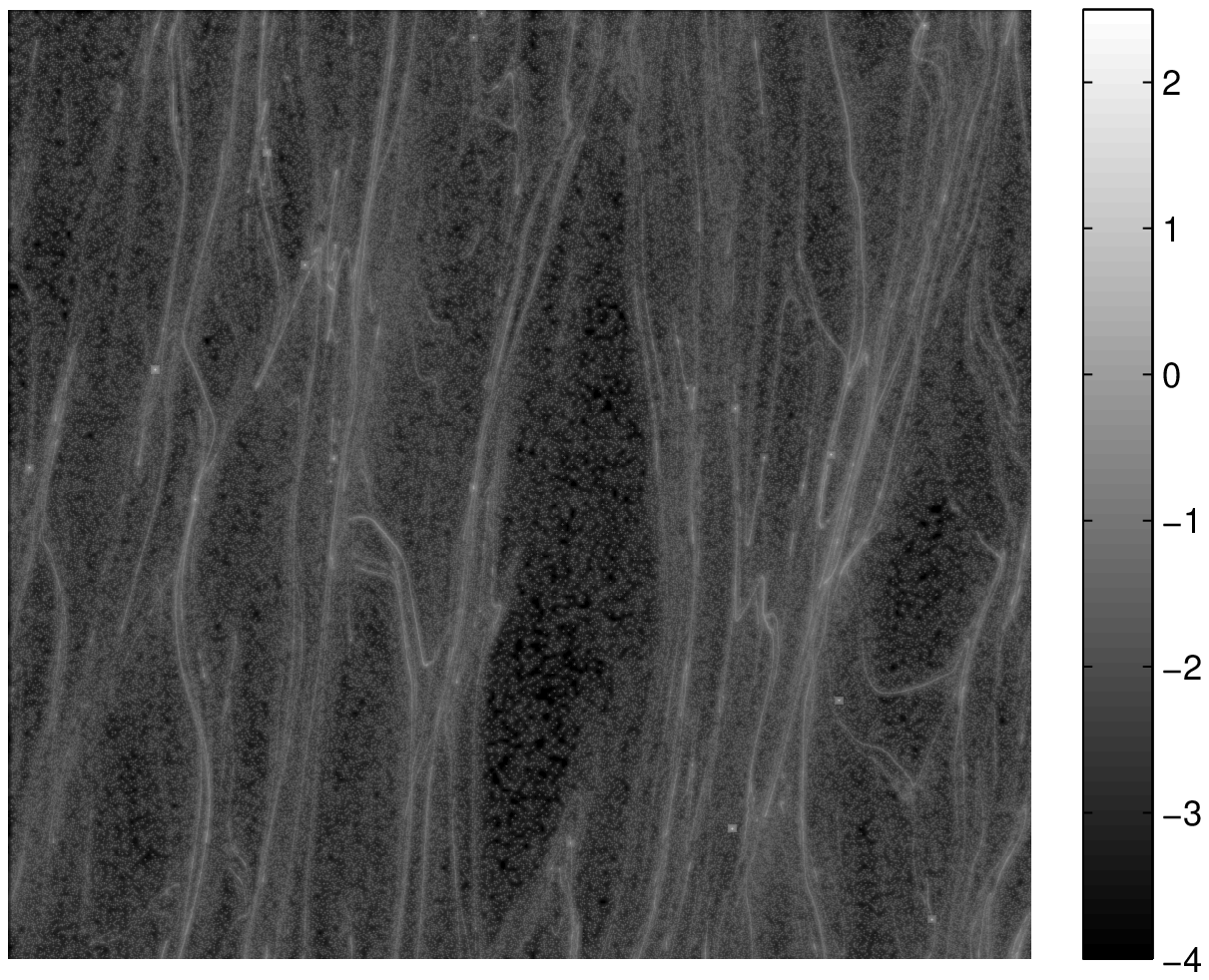}
\caption{Logarithmic surface density of the gas (left) and particles
(right) in the quasi-steady state in the run starting with a lower
surface density of particles, $\Sigma_{p,0} = 10^{-3}\Sigma_0$, at
$t_c = 10\Omega^{-1}$. The dust particles are preferentially
collected in the over-densities (crests) of density waves formed due
to the gravitational instability in the gas. Because of the low dust
mass, the gas and particle dynamics are largely unaffected by the
particle self-gravity and by the back-reaction via drag force.}
\label{tc10_lm}
\end{figure*}

\begin{figure*}
\includegraphics[width = 0.49\textwidth]{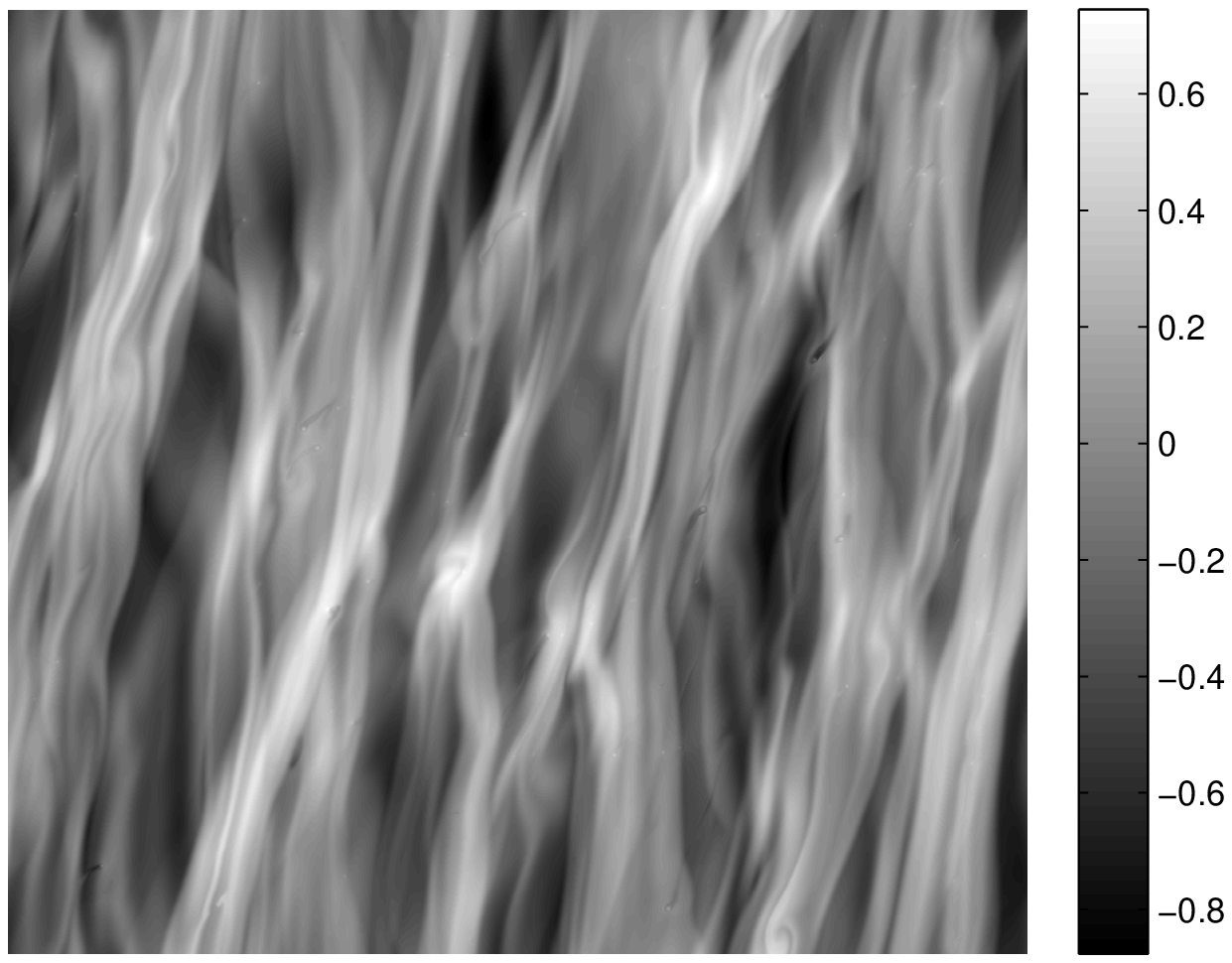}
\includegraphics[width = 0.49\textwidth]{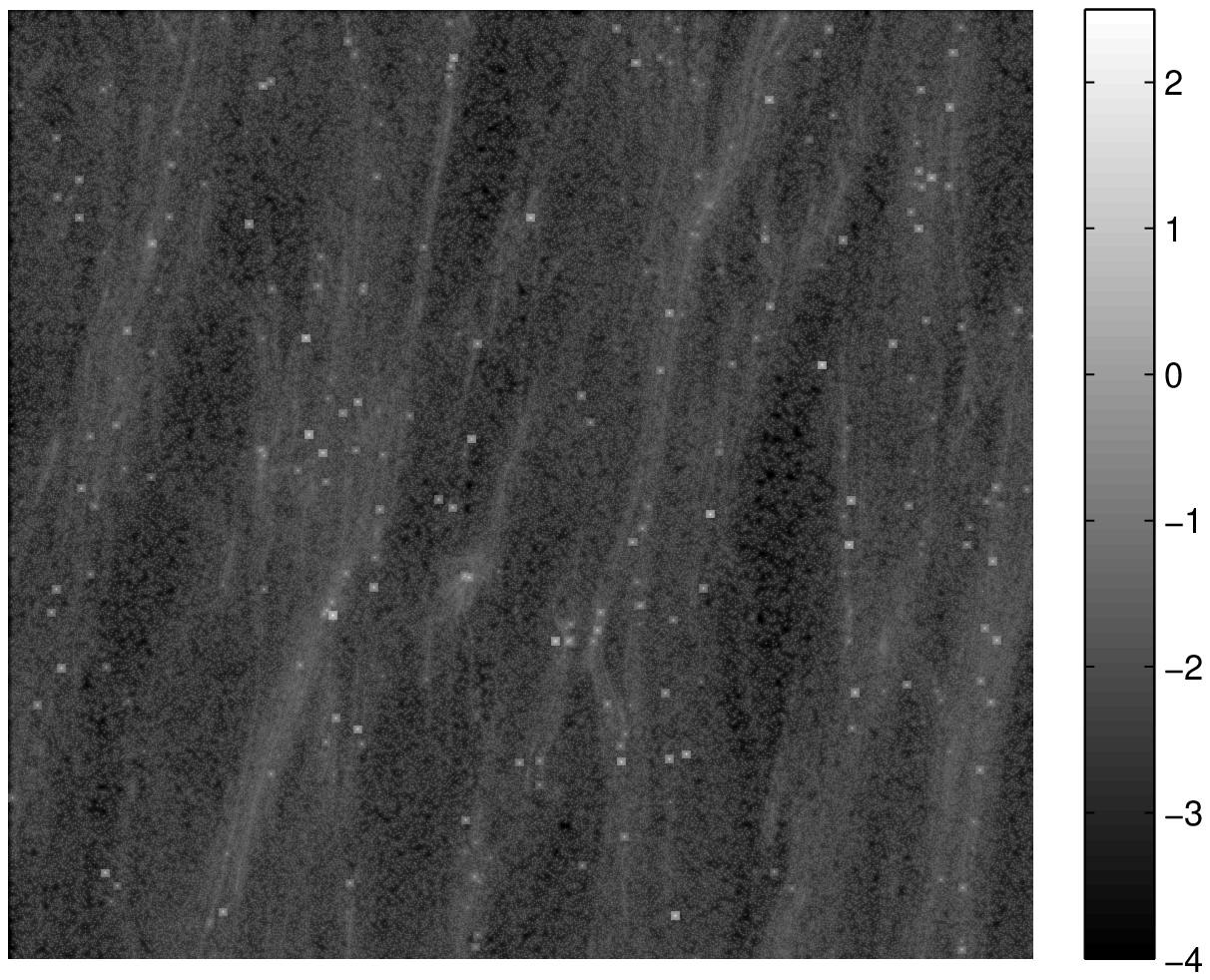}
\caption{As in Fig. 1, but for the standard simulation starting with
$\Sigma_{p,0}=10^{-2}\Sigma_0$ and $t_c = 10\Omega^{-1}$. Particles
are trapped in the over-densities created by density waves in the
gas. Due to the particles' gravitational interaction with each
other, some of these trapped particle groups, which happen to reach
large enough concentrations, collapse and form very dense bound
clumps (white dots). Part of these clumps have so high densities
that their back-reaction on the gas results in the wake of the
clumps' motion appearing in the gas.}
  \label{tc10}
\end{figure*}

\begin{figure*}
\includegraphics[width = 0.49\textwidth]{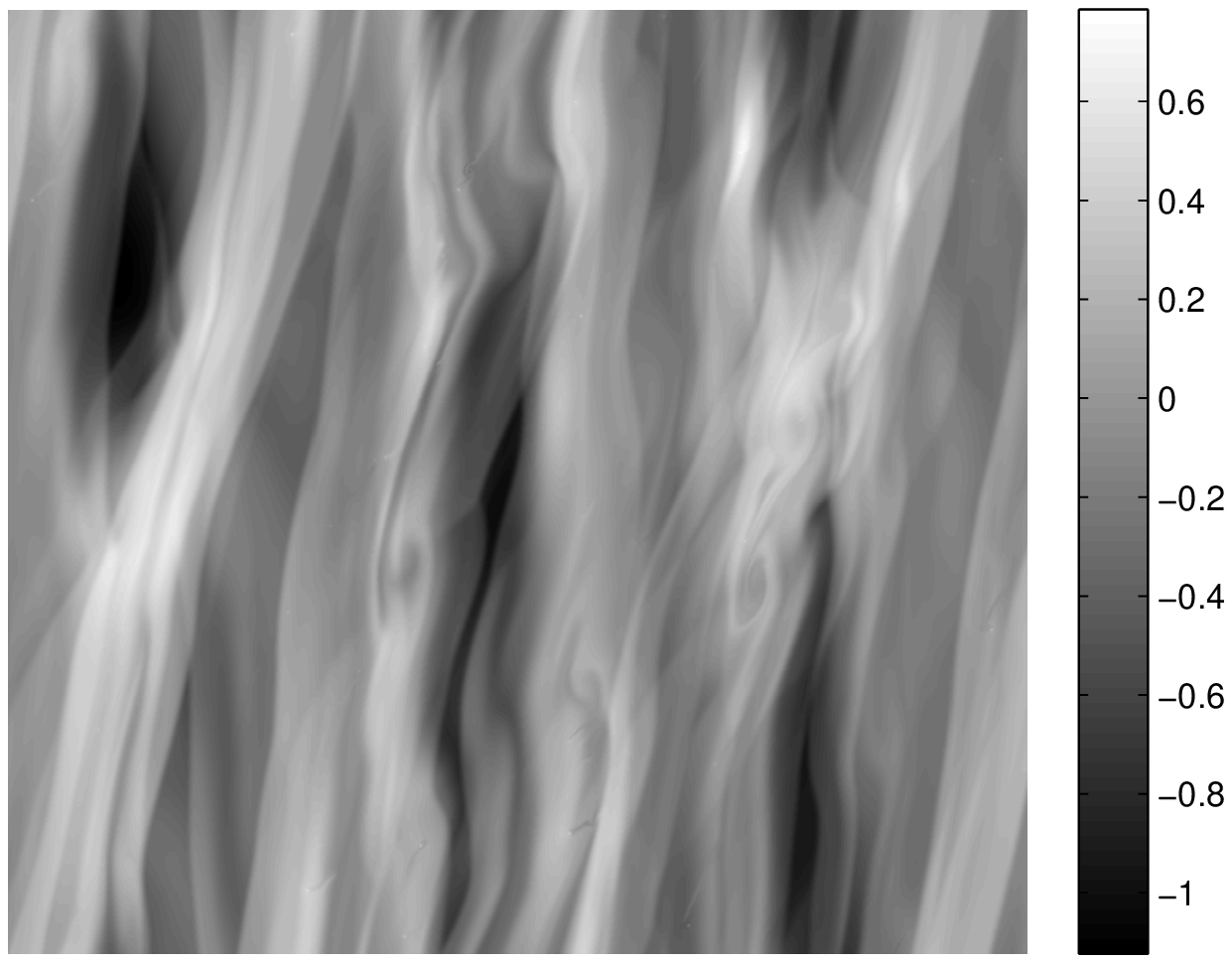}
\includegraphics[width = 0.49\textwidth]{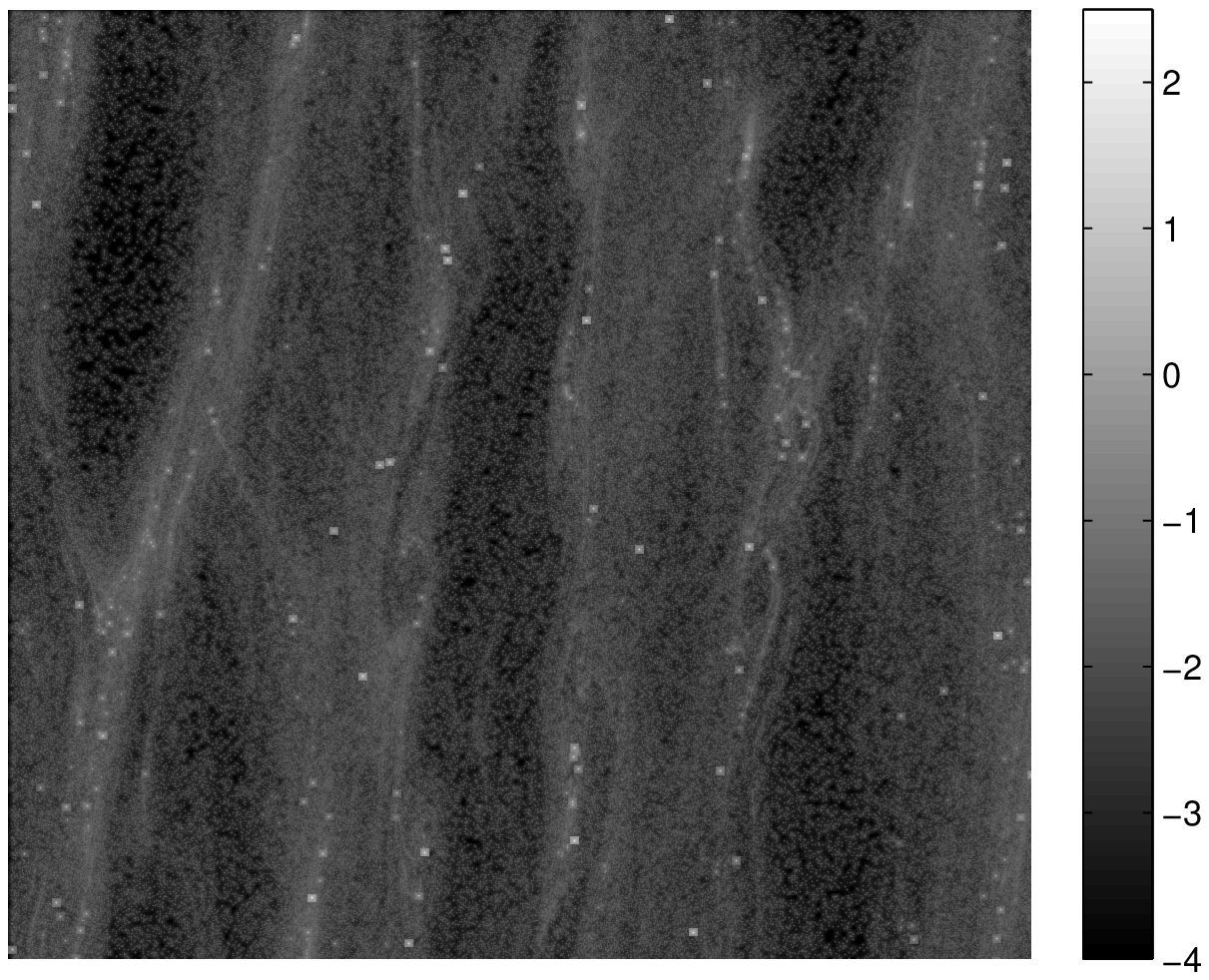}
\caption{Same as in Fig. 2, but for $t_c =
20\Omega^{-1}$.}\label{tc20}
\end{figure*}

\begin{figure*}
\includegraphics[width = 0.49\textwidth]{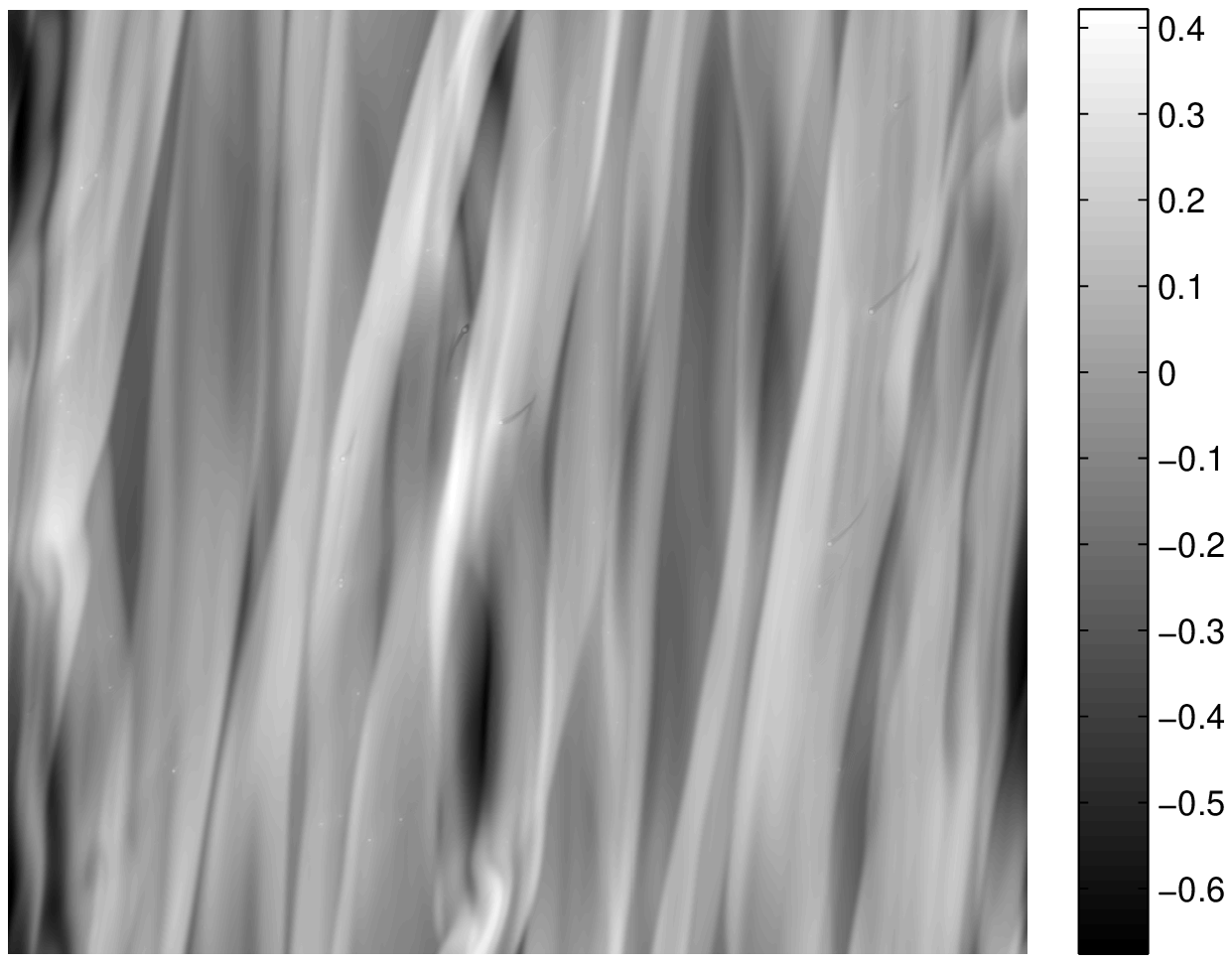}
\includegraphics[width = 0.49\textwidth]{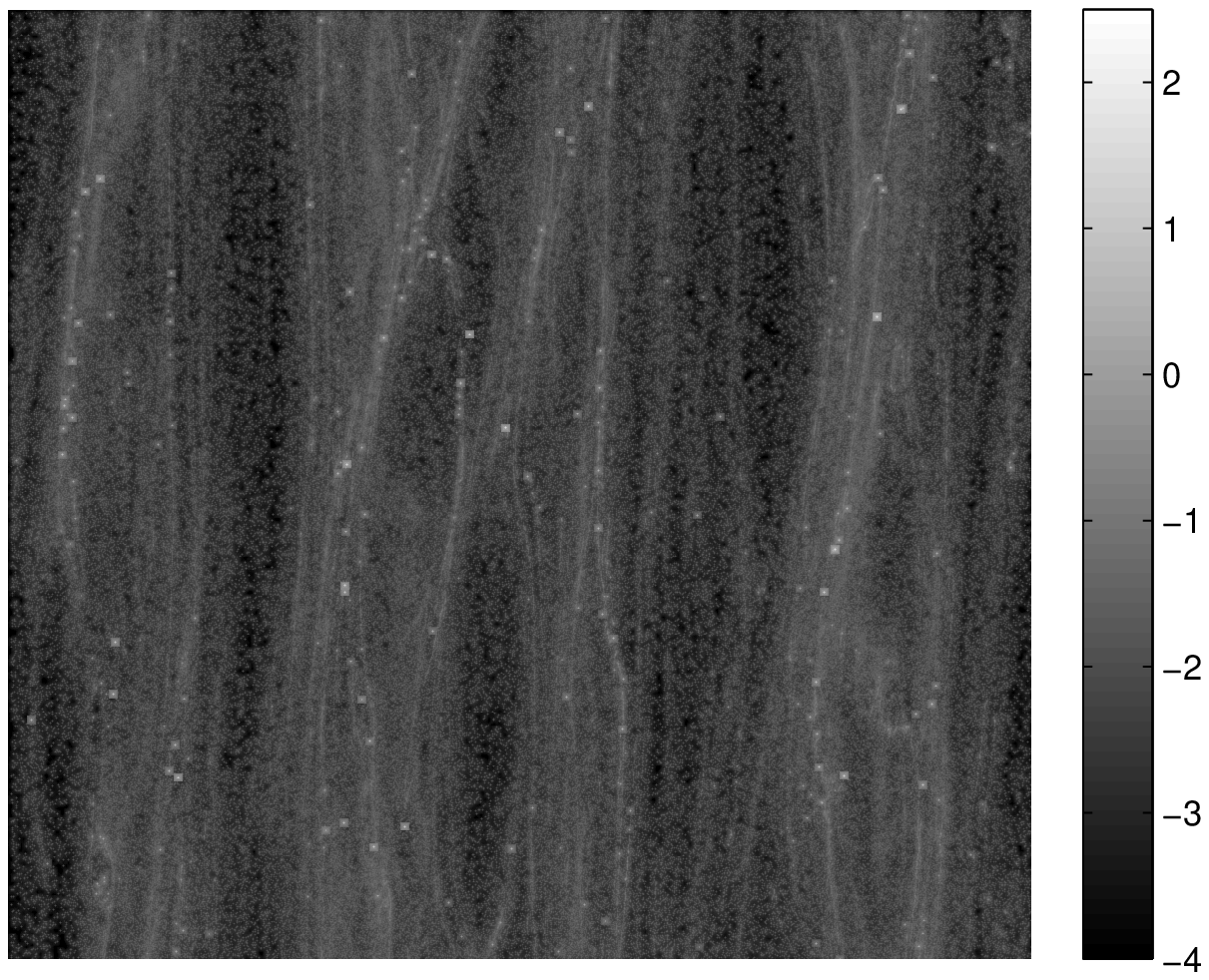}
\caption{Same as in Fig. 2, but for $t_c = 40\Omega^{-1}$. Wakes of
three bound clumps of particles are discernible in the gas density
map.} \label{tc40}
\end{figure*}

\begin{figure*}
\includegraphics[width = 0.49\textwidth]{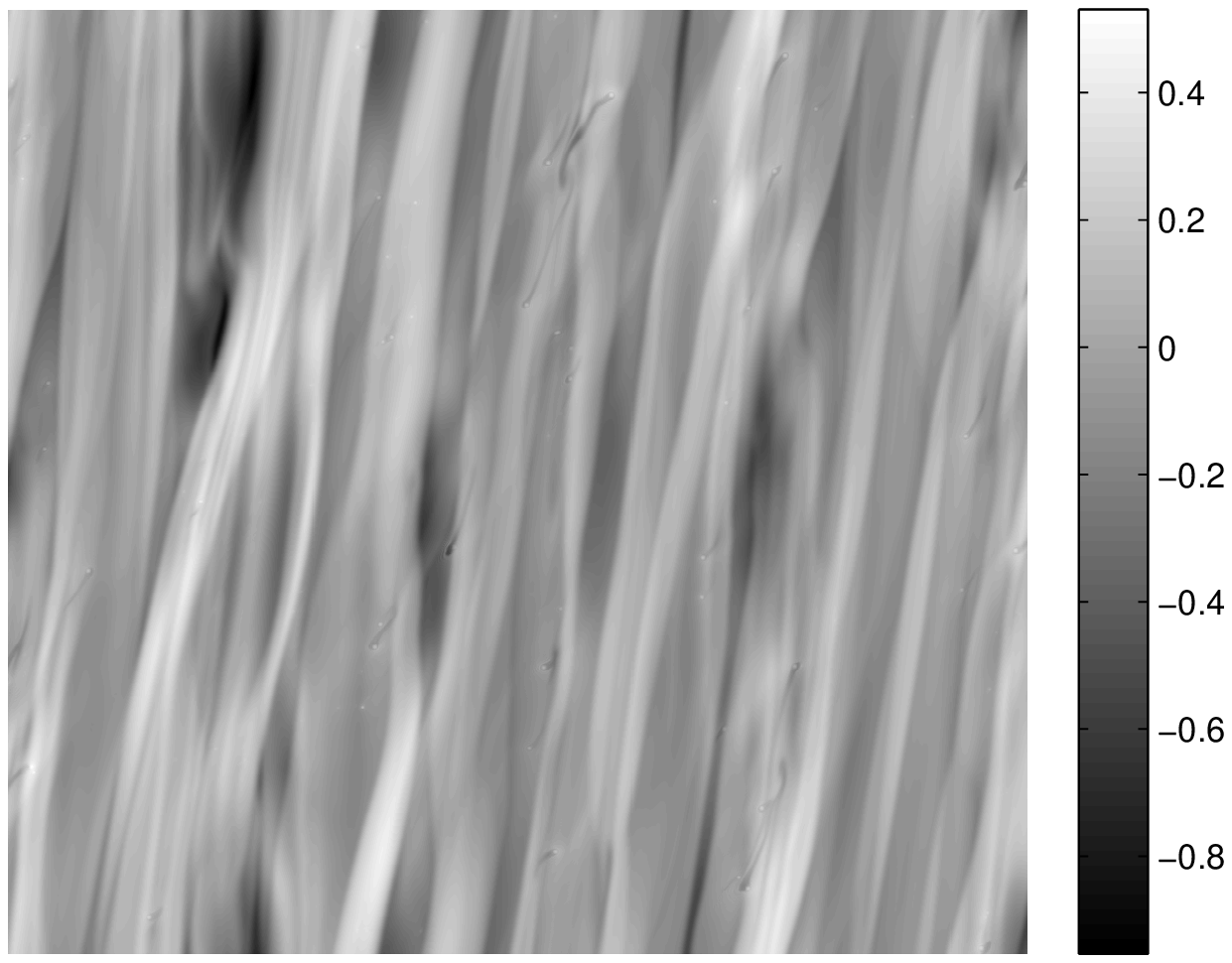}
\includegraphics[width = 0.49\textwidth]{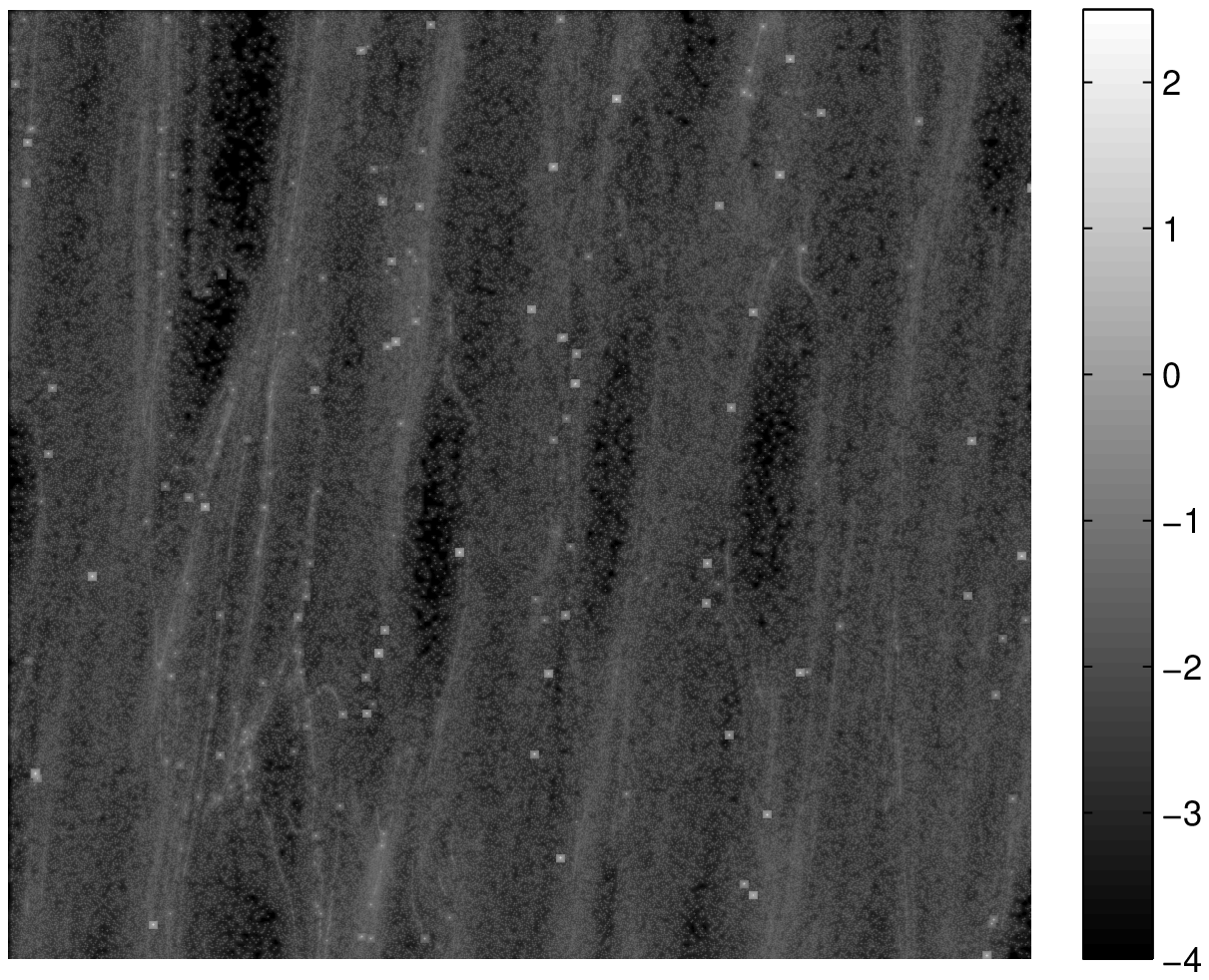}
\caption{Same as in Fig. 2, but for $t_c = 80\Omega^{-1}$. Wakes of
several bound clumps of particles are clearly visible in the gas
density map.} \label{tc80}
\end{figure*}

In this study, as noted above, we also include the self-gravity of
particles and back-reaction of the particles' drag force on the
evolution of the gas, the processes omitted in Paper I, due to the
massless test particles adopted by that study. Here we include the
back-reaction for both values of initial dust-to-gas mass ratios,
0.01 and 0.001, considered. Figures \ref{tc10_lm} - \ref{tc80} show
the surface density of the gas and particles (for all species, i.e.,
friction times) in the quasi-steady gravitoturbulent state at the
end of each simulation performed. Generally, the evolution of the
gas appears to be almost identical to the results presented in Paper
I, where the particles do not affect the gas, however, close
examination of the $\Sigma_{p,0}/\Sigma_0=10^{-2}$ runs shows that
when the particle density reaches extremely high concentrations, the
wakes of large clouds of particles are visible in the gas. These
wakes appear to become more pronounced as the cooling time
increases, possibly due to the density waves having less robust
structure in the simulations with longer cooling times. However, in
none of the runs do these features appear to have a significant
influence on the overall evolution of the gas.
\begin{figure*}
\includegraphics[width=0.49\textwidth]{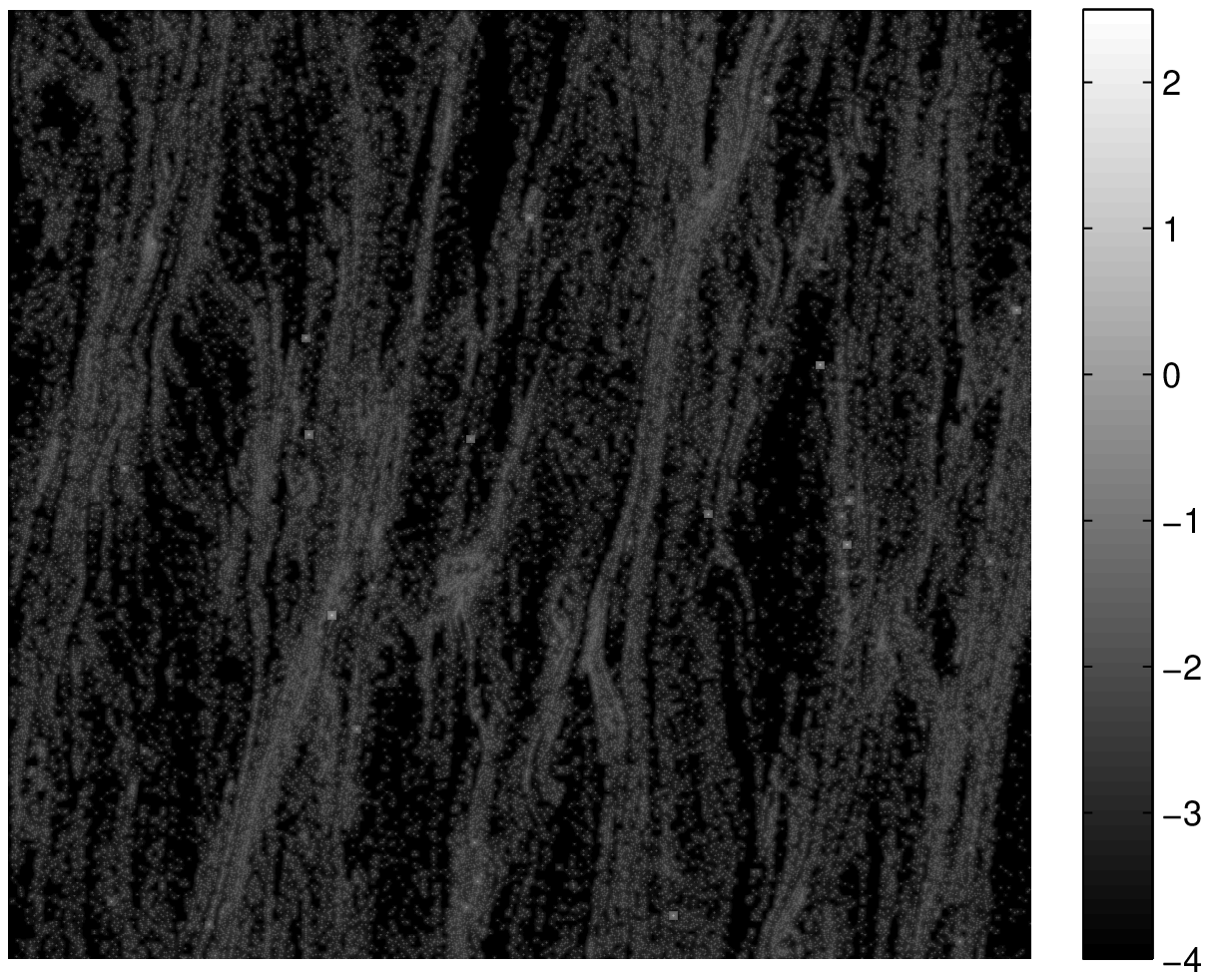}
\includegraphics[width=0.49\textwidth]{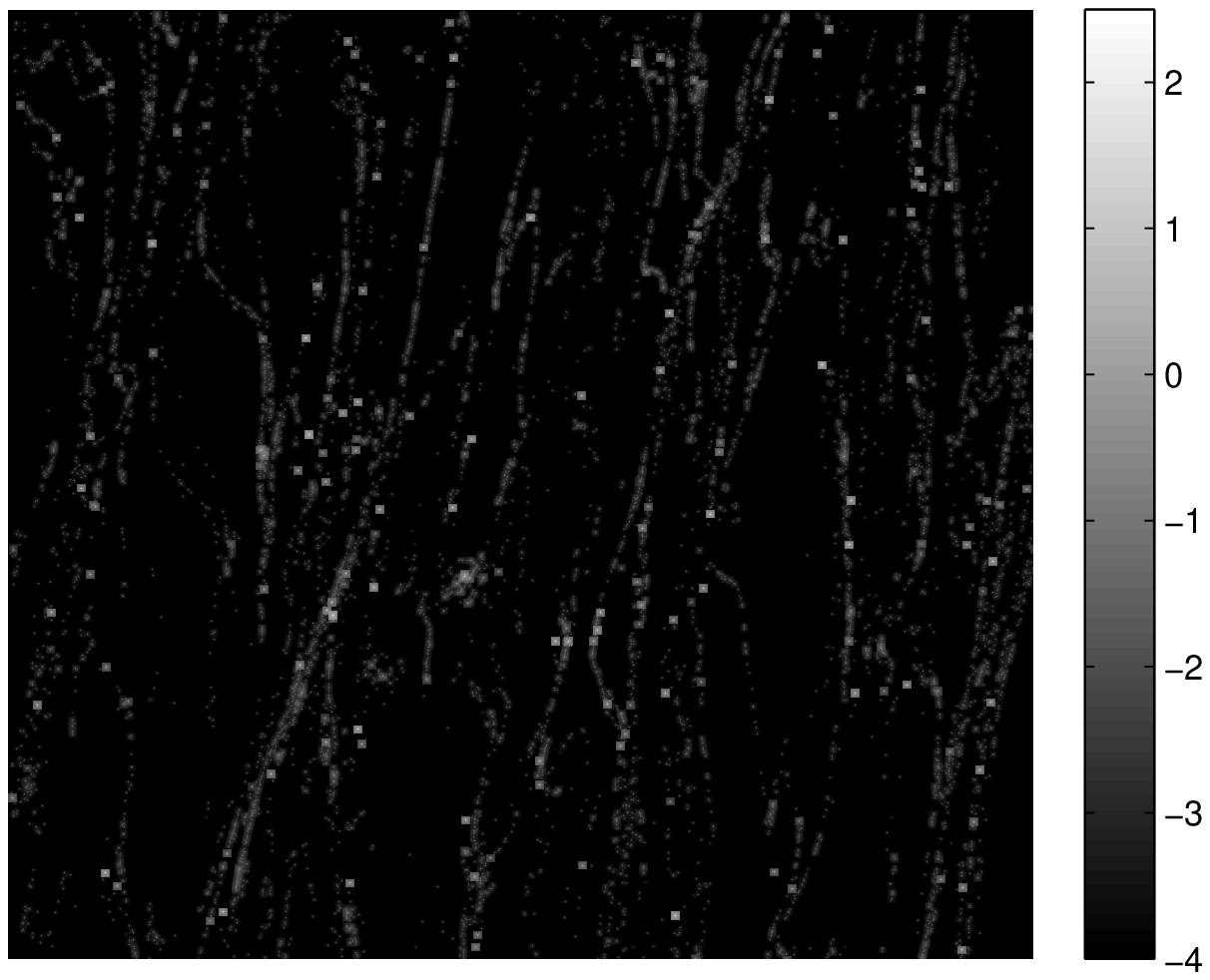}
\includegraphics[width=0.49\textwidth]{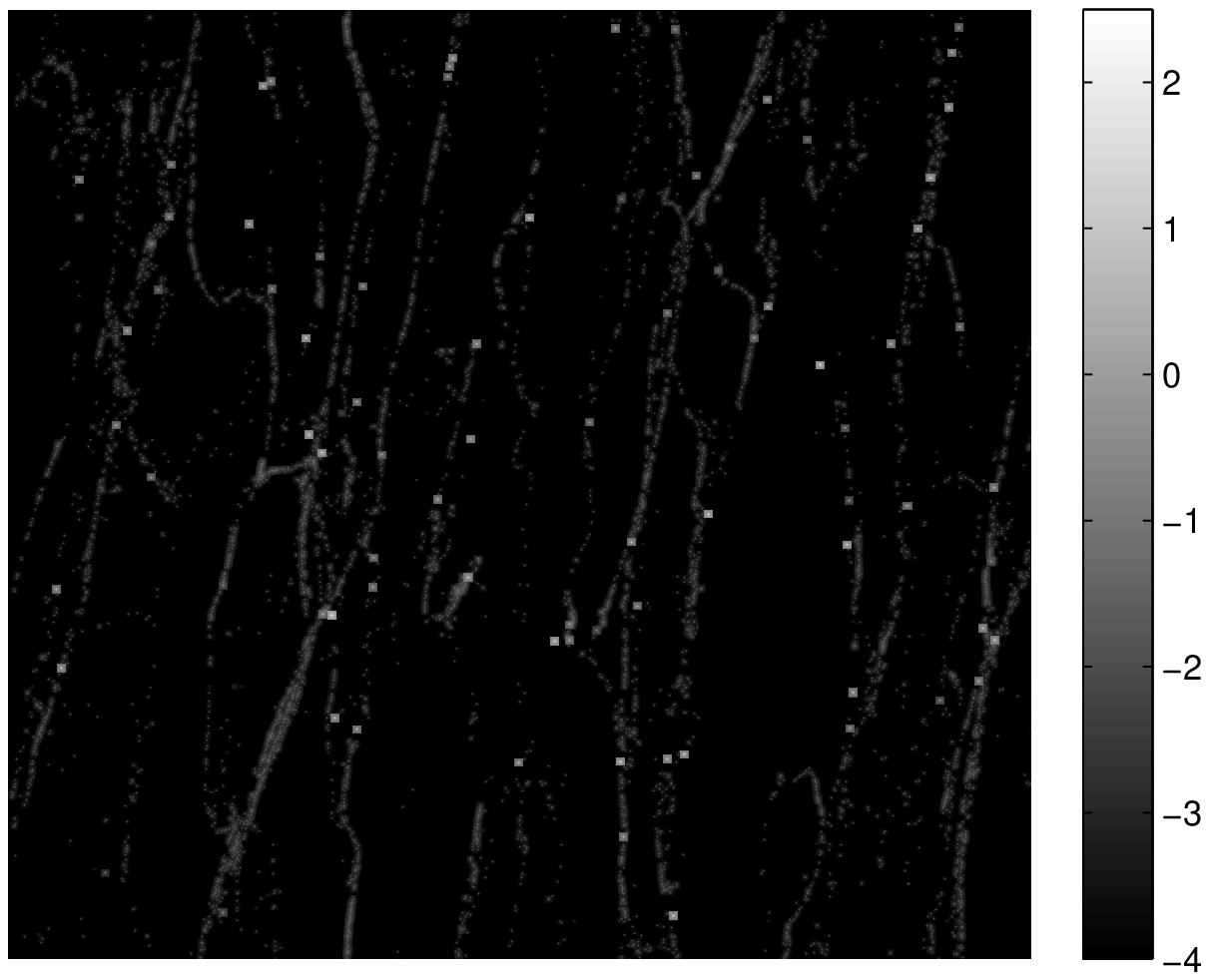}
\includegraphics[width=0.49\textwidth]{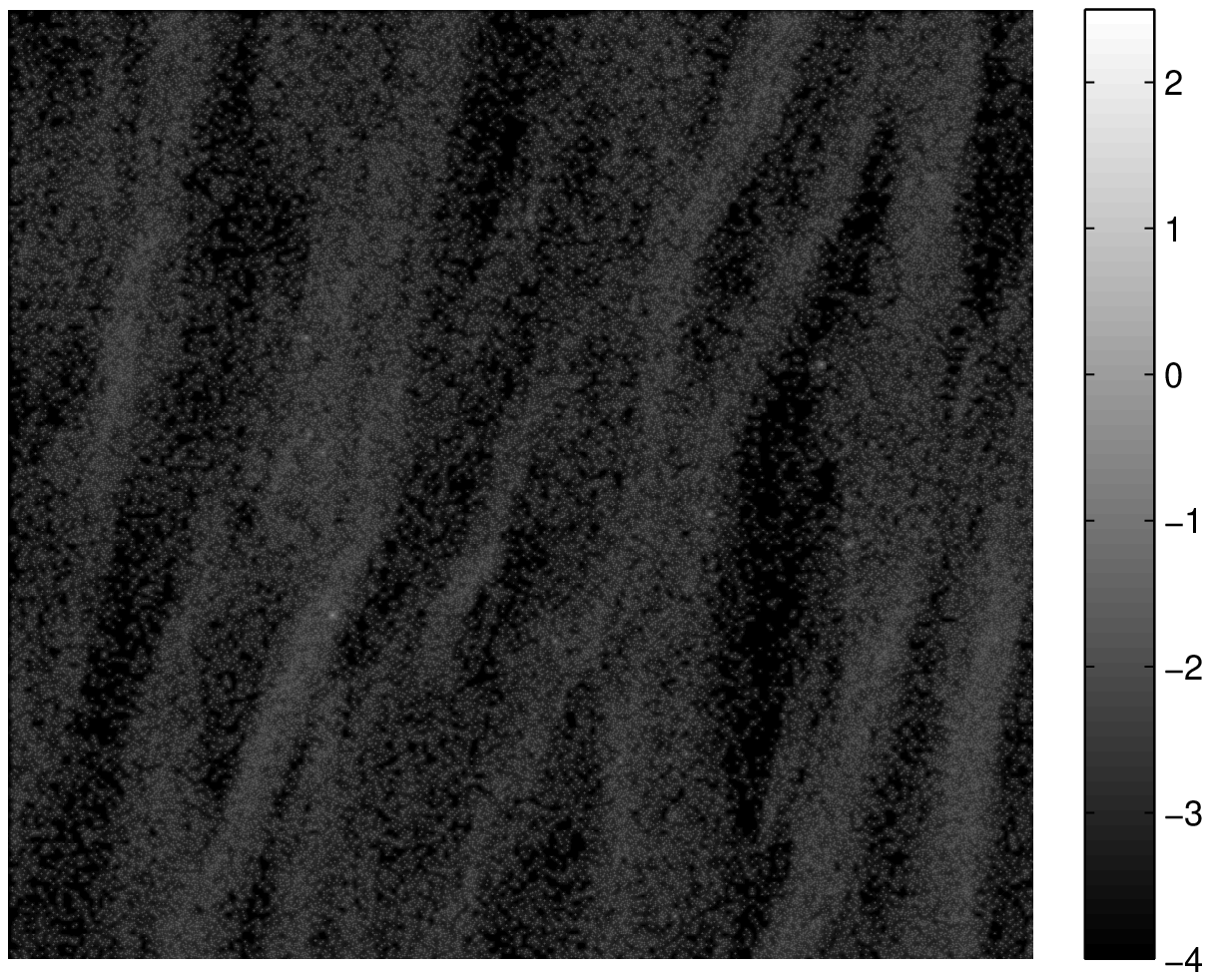}
\includegraphics[width=0.49\textwidth]{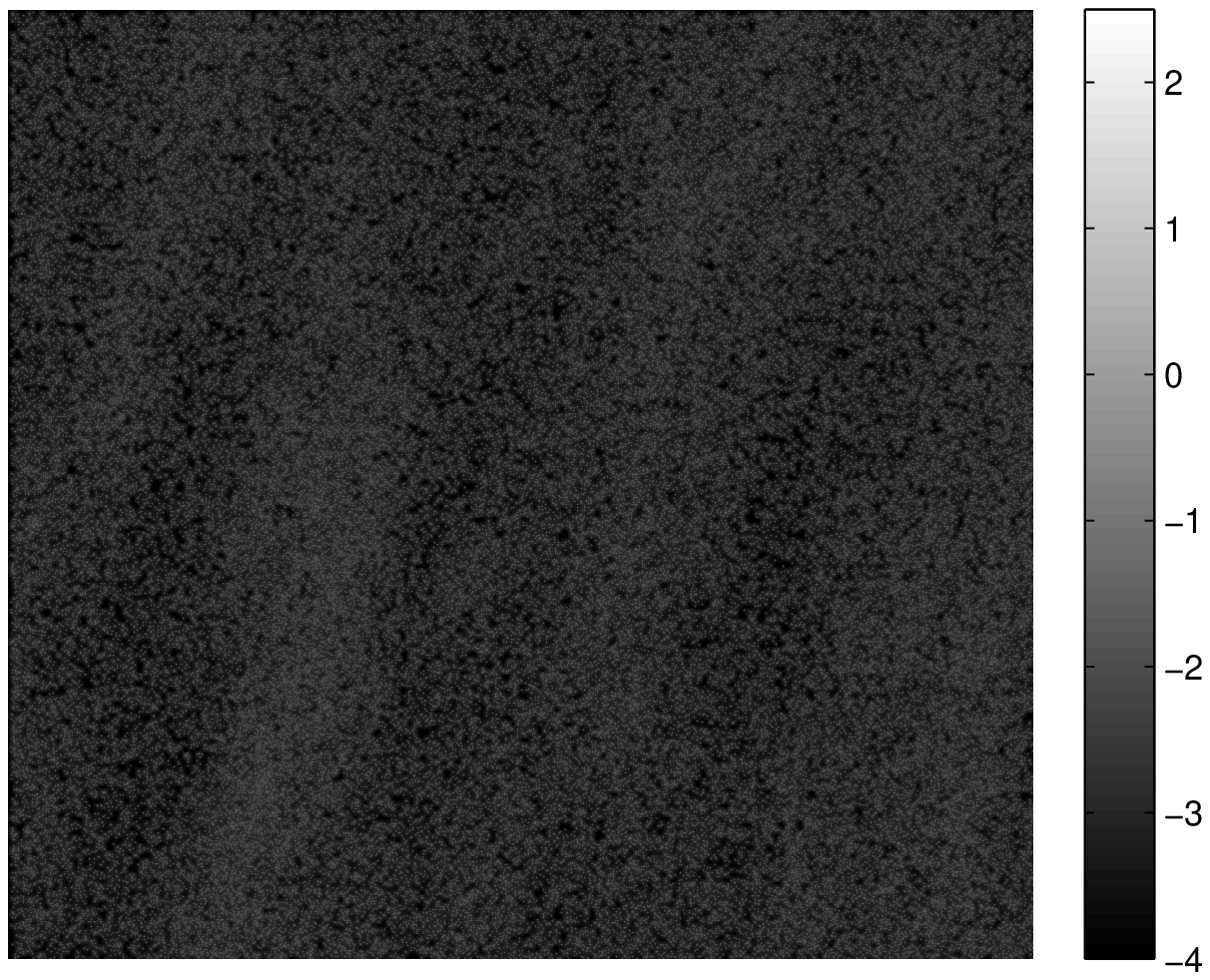}
\caption{Logarithmic surface densities of the dust particles with
different friction times in the run with
$\Sigma_{p,0}=10^{-2}\Sigma_0$ and $t_c = 10\Omega^{-1}$ (Fig. 2) at
$t=40T$, when the disc is already in a quasi-steady state, or after
30 orbital times since the drag force between the gas and particles
has been turned on. The total particle density field shown in Fig. 2
has been decomposed according to particle friction times $\tau_f =
[0.01,0.1,1,10,100]\Omega^{-1}$, so that each panel shows the
surface density of particles with a specific friction time from the
lowest in upper left panel to the highest in the bottom middle
panel. In the density maps of the $\tau_f = 0.1\Omega^{-1}$ (top
right) and $\tau_f = 1.0\Omega^{-1}$ (middle left) particles, we
clearly see a number of highly dense structures (white dots), which
are gravitationally bound.} \label{Stopping_Times}
\end{figure*}

\subsection{Particle concentration}

The particles, which initially have zero velocities (relative to the
Keplerian flow) and random locations within the simulation domain,
are not evolved until the gas has undergone an initial burst phase
of gravitational instability and settled into a quasi-steady state.
Once the gas has reached this state, we release the dust particles.
For the $t_c = 10\Omega^{-1}$ and $20\Omega^{-1}$ runs, the
particles are introduced at a time $t_{par} = 10T$, while for the
$t_c = 40\Omega^{-1}$ and $80\Omega^{-1}$ runs they are introduced
at $t_{par} = 20T$ and $t_{par} = 30T$ respectively. At this point,
we switch on the drag force between the gas and the dust particles,
evolving the system for a further 5 orbits in each case until the
dust particles have begun to trace the structure of the gas.
Finally, we introduce the self-gravity of the particles, evolving
the system with both the drag force and particle self-gravity for
another 25 orbits.

Once the particles are introduced, they are drawn to local pressure
maxima associated with the density waves in the gas. Figure
\ref{tc10_lm} shows the logarithmic surface densities of both the
gas and dust grains once the system has reached a quasi-steady state
at $t = 40T$ in the run with $t_c = 10\Omega^{-1}$ and
$\Sigma_{p,0}=10^{-3}\Sigma_0$. Comparing the gaseous and dust
components of the simulation, we see the same correlation between
the density enhancements in the gas and over-densities in the
particles, as also observed in Paper I. The degree to which the
presence of spiral density waves affect the particle concentration
depends strongly on the friction time of the particles. As found in
Paper I, the smaller particles with friction times $\tau_f =
[0.01,0.1,1.0]\Omega^{-1}$, tightly map the structure which forms
due to gravitational instabilities in the gas, whilst the larger
particles are not as affected by the drag force and their evolution
is not significantly altered by the structures and motion of the
gas. Comparing this low particle density case with both the higher
particle density case in Fig. \ref{tc10} and the previous massless
particle case considered in Paper I, we see that particle
self-gravity for low mass particles does not have a significant
effect on their evolution. The evolution of particles in this case
is qualitatively the same as that of the massless ones.

Figures \ref{tc10} - \ref{tc80} show the particle surface density in
the simulations with $\Sigma_{p,0}=10^{-2}\Sigma_0$ and different
cooling times. As outlined in Paper I, the cooling time imposed on
the disc is related to the location of the simulation domain within
the disc \citep{Rafikov2005,RiceArm2009,Clarke2009}, with the range
of cooling times considered spanning the radial range 20-60AU for
typical disc parameters. Typically, this radial interval corresponds
to saturated quasi-steady self-gravitating state characterized by
$Q\sim 1$ in discs; at larger radii fragmentation is expected,
whereas at smaller radii the effect of self-gravity becomes weaker,
i.e., $Q$ increases \cite[see e.g.,][]{Boley2006,Clarke2009}. It is
seen in Figs. \ref{tc10} - \ref{tc80} that at all cooling times
considered, the accumulation of particles within gas over-densities
created by spiral density waves leads to the formation of very high
density clouds of particles there. These particle clouds form
exclusively as a result of the inclusion of the particle
self-gravity term. Once the particle surface density in the clouds
reaches a certain value, typically equal to the gas local density,
gravitational instabilities set in in the solid component of the
disc, causing these filaments of particles that form within spiral
density wave crests to contract into several highly dense,
gravitationally bound objects (white dots on the particle density
maps in Figs. 2-5). Figure \ref{Stopping_Times} shows the surface
density of the particles at the end of the
$\Sigma_{p,0}=10^{-2}\Sigma_0$, $t_c = 10\Omega^{-1}$ simulation
decomposed into separate friction times. Here we see that the clouds
are primarily composed of intermediate size particles with friction
times $\tau_f = [0.1,1.0]\Omega^{-1}$. Particles with these friction
times tend to most efficiently concentrate into the density waves
(see also Paper I) and make up the bulk of the mass that is
contained in the clumps formed due to the particles' self-gravity,
for example, over 90\% of the mass in the most massive particle
cloud in this simulation is made up of particles with these friction
times.

\begin{figure}
\includegraphics[width = 0.5\textwidth]{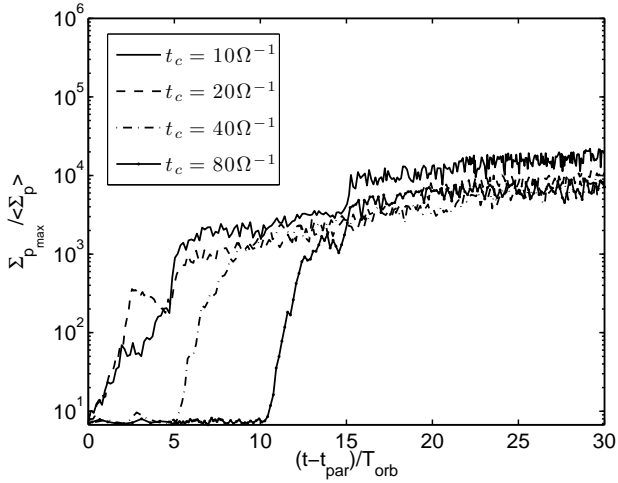}
\caption{Maximum surface density of the dust particles within the
domain as a function of time for each cooling time. There is little
correlation between cooling time and the maximum surface density
reached, implying the particle self-gravity can produce significant
density enhancements in the solid component, even for longer cooling
times} \label{Sigma_max_t}
\end{figure}

The inclusion of self-gravity in the evolution of the dust particles
causes their surface density to reach values $\sim 10$ times higher
than those in the absence of particle self-gravity. Figure
\ref{Sigma_max_t} plots the maximum surface density of the particles
over the simulation domain as a function of time at different
cooling times. Although for longer cooling times it takes
accordingly longer for the particle surface density to reach a
maximum value, this maximum appears to be almost independent of the
cooling time, instead being determined by the strength of particle
self-gravity (i.e., the total mass of particles in the simulation
domain).

As explained in Paper I, if we specify an accretion rate, based on
the analytic description of a quasi-steady self-gravitating disc by
\citep{Clarke2009,RiceArm2009}, we can obtain a cooling time-radius
relation, allowing us to determine at what radius the cooling times
modelled here lie in a realistic disc. In these papers, the surface
density profile of a quasi-steady self-gravitating disc is
calculated, therefore allowing us to obtain a characteristic surface
density, $\Sigma_0$, for each of our simulations as a function of
the fiducial radius $r_0$. Also, recalling the physical length of
the box, $L_x =L_y=80G\Sigma_0/\Omega^2$, we obtain the total mass
of gas in the domain,
\[
M_g=\Sigma_0L_xL_y=80^2\frac{G^2\Sigma_0^3}{\Omega^4},
\]
or making use of the relation between the mass accretion rate,
$\dot{M}$, and the $\alpha$ parameter, $\dot{M}=3\alpha c_{s0}^3/GQ$
for a self-gravitating disc in a steady state \citep{Clarke2009},
\[
M_g=\frac{80^2}{9^{2/3}\pi^4
}\cdot\frac{1}{\Sigma_0}\left(\frac{\dot{M}}{\alpha
Q^2G^{1/2}}\right)^{4/3}.
\]
Adopting the values of the quantities in this expression at radii
larger than $20$AU based on the self-gravitating disc structure
models of \citet{RiceArm2009}, $\Sigma_0=20~{\rm g~cm^{-2}}$, $Q=1$,
$\alpha=0.04$ and $\dot{M}=10^{-7}M_\odot~{\rm yr}^{-1}$, we get
$M_g=0.02M_\odot$, so the total dust mass enclosed within the
simulation domain is $2\times 10^{-4}M_g$. Dividing the dust mass by
the number of particles $5\times 10^5$, we find the physical mass of
each numerical super-particle, $1.33\times 10^{-4}M_{Earth}$. By
identifying the regions of highest particle concentrations and
calculating the velocities of each particle therein relative to the
velocities of its neighbour particles (defined here to be particles
within a grid cell distance $\Delta r = \Delta x$) and comparing
with their mutual gravitational potential energy, the boundness of
each particle aggregate can be evaluated.

\begin{figure}
\includegraphics[width = 0.5\textwidth]{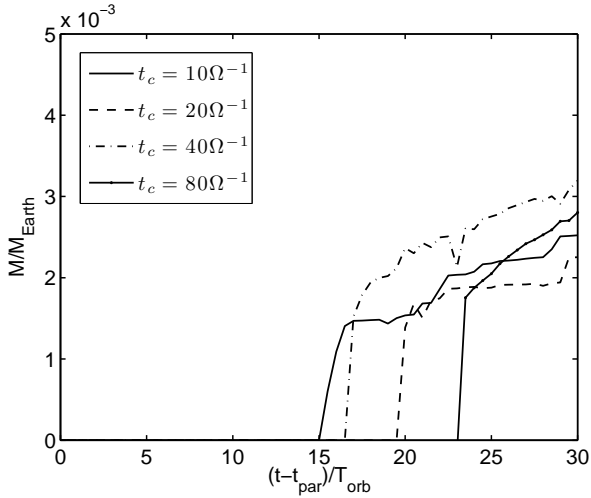}
\caption{Mass of the largest gravitationally bound collection of
particles in the domain as a function of time at each cooling time.}
\label{Max_mass_t}
\end{figure}

Figure \ref{Max_mass_t} shows the mass of the largest
gravitationally bound aggregate of particles in the domain as a
function of time. It is seen that the mass of clumps that form does
not really correlate with cooling time (and by extension with radius
within the disc). Any increase in particle concentration which
occurs in the case of lower cooling time tends to be offset by the
larger surface density of particles at inner radii where cooling
times are relatively long.

\begin{figure}
\includegraphics[width = 0.5\textwidth]{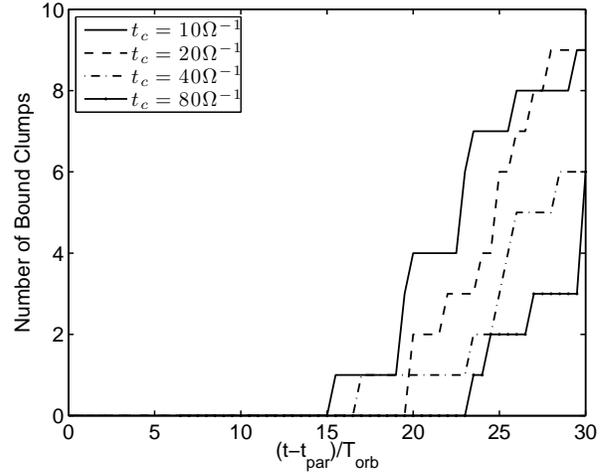}
\caption{Number of gravitationally bound concentrations of particles
in the domain as a function of time at each cooling time.}
\label{Bound_Clumps}
\end{figure}
\begin{figure}
\includegraphics[width = 0.5\textwidth]{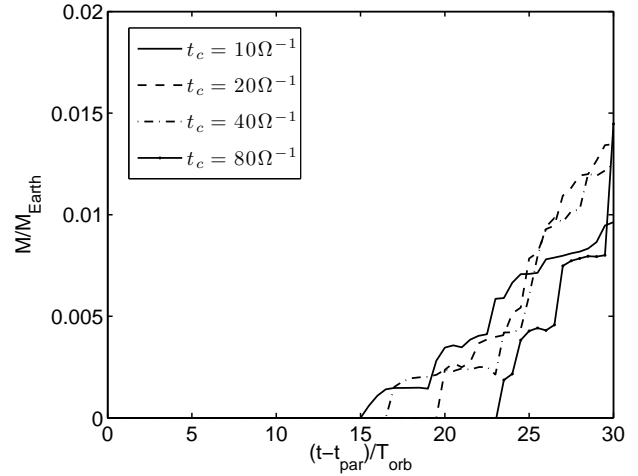} \caption{Total mass
of gravitationally bound concentrations of particles in the domain
as a function of time at each cooling time.} \label{Bound_Mass}
\end{figure}

In practice, the particles will also undergo collisions as a result
of being so densely packed in a given bound clump, potentially
leading to their destruction and inhibiting planetesimal formation.
As we showed in our previous studies \citep[][Paper I]{Rice2006},
the velocity dispersion of particles in self-gravitating discs is
usually comparable to the gas sound speed, $\sim
100-1000{\rm~m~s^{-1}}$, which is much larger than the break-up
collisional speed threshold \citep[$1-10~{\rm m~
s^{-1}}$][]{Benz2000, BlumWurm2008}. But as shown in this paper, the
main factor that holds the $\tau_f \sim 1.0\Omega^{-1}$
particles\footnote{This friction time, at which particle
concentration in density waves is most efficient, corresponds to
particle sizes 3-10 cm in the radial range 20-60AU where the disc
resides in the quasi-steady self-gravitating state with $Q\sim 1$
\citep[see][Paper I]{Clarke2009,RiceArm2009}. For comparison, in
\citet{Rice2006}, where the disc is confined to a maximum radius
25AU, the optimally trapping friction time $\tau_f \sim
1.0\Omega^{-1}$ corresponds to metre-sized particles.} together into
bound clumps, despite such high velocities, is their self-gravity
which is effective due to sufficiently high particle concentrations
achieved in gaseous density waves \cite[see also][]{Rice2006}. These
clumps are bound in the sense that the velocity dispersion of their
constituent particles is smaller than the escape velocity of
self-gravitational binding energy of the mass enclosed within the
clump. Of course, inside individual clump, particles can collide and
even undergo fragmentation because of their large velocities. So, it
is likely that each such clump may represent a number of small
fragments rather than a single object. However, it is shown in Paper
I that, in density waves, particle velocities tend to align due to
drag force and therefore their relative velocities turn out to be
smaller, possibly preventing collisional destruction. The resolution
limits of our simulations do not allow us to follow the subsequent
evolution of each bound clump, since it is smaller than the grid
cell and probably continues to shrink even further due to its own
gravity as a result of dissipation of particle velocity dispersion
by drag force \citep[see][]{Lyra2009}, so we only determine clump
masses (i.e., how many particles are trapped in a given clump), like
other related simulations addressing particle capture in gaseous
structures in discs
\citep[][]{Lyra2008b,Lyra2009,Johansen2007,Johansen2011}. One of the
main results of this study is to show that such gravitationally
bound, long-lived clumps can do form when particle self-gravity is
included. With a more detailed analysis, which takes into account
the effects of collisions between particles, a clearer picture of
the evolution of these `rubble piles' can be built. By replacing the
accumulations of particles presented here with a single massive
particle, and modelling the accretion of gas from the disc onto the
growing planet, the late stages of planet formation could be
studied.

Figures \ref{Bound_Clumps} and \ref{Bound_Mass} also show the number
of particle clumps in the domain, which are gravitationally bound,
next to the total mass of particles contained in gravitationally
bound structures with time. From these figures, we see that at
shorter cooling times clumps tend to form over shorter time
intervals, and the runs with a shorter cooling time produce more
gravitationally bound structures over the time period modelled. The
total mass contained in bound structures is nearly independent of
the cooling time, for similar reasons mentioned for Fig.
\ref{Max_mass_t}. Although the simulations adopting a shorter
cooling times tend to produce more clumps, and concentrate a higher
fraction of the total particle mass available in the domain into
bound clumps, the higher particle surface densities associated with
the inner radii result in the total mass contained in bound
structures being still somewhat higher.

\section{Summary and discussions}

The work presented in this paper expands on the findings of Paper I,
presenting a series of simulations modelling the evolution of dust
particles in the presence of spiral density waves occurring as a
result of gravitational instabilities in the disc. In particular,
our study focuses on the effect that the gravitational interaction
between massive dust particles, or their self-gravity, has on their
evolution, expanding on the massless `tracer' particles adopted
previously by also taking into account the particle back-reaction on
the gaseous component via drag force. A general picture of the
evolution of the dust particles remains unchanged, they evolve
through a quasi-steady gaseous density structures associated with
density waves produced by a combined effect of disc self-gravity and
cooling. Particles accumulate in high-density/pressure regions of
spiral density waves, producing as a result significant
over-densities in the solid component of the disc, with the
magnitude of the particle density enhancement depending on both the
cooling time of the gas and the friction time of the particles.

The inclusion of the particles' self-gravity can have several
significant effects on the evolution of the disc. The intensity of
these effects depends on the total mass of particles in a given
simulation. If this mass is low (i.e., the dust-to-gas mass ratio
$\ll 0.01$) particles' self-gravity has no significant effects on
the evolution of either the gas or solid component of the disc. For
more canonical dust-to-gas mass ratios ($\sim 0.01$), particle
self-gravity causes the particle aggregates, which are trapped in
the crests of spiral density waves, to contract further. If such
particle concentrations reach high enough densities, gravitational
interactions among particles inside become sufficiently large to
cause local collapse of the solid component of the disc, leading to
the formation of gravitationally bound structures within the disc.
This picture, obtained within the local shearing sheet approach
permitting higher numerical resolution, is consistent with the
results of analogous global simulations by \cite{Rice2006} of
particle dynamics in self-gravitating discs which also take into
account self-gravity of dust component. Assuming typical disc
parameters, however, predicts masses for these structures to be
comparable to those of very large planetesimals, with the most
massive structures identified, comparable to the mass of large
asteroids and dwarf planets, potentially providing seed objects for
the growth of terrestrial planets and the cores of gas giant planets
as outlined in the calculations of \citet{Pollack1996}. These
structures are robust enough to survive in the disc, even after the
`parent' spiral density wave in which they formed has been sheared
out and the remainder of the solid particles have diffused back into
the disc. Interestingly, the physical mass of these structures is
only weakly related to the cooling time of the disc. Although at
short cooling times particles get trapped within density waves more
efficiently, resulting in structures forming faster and usually
accounting for a larger fraction of the particles, for a given disc,
the lower surface density of material present at the larger radii
associated with these shorter cooling times tends to offset this.
This suggests that density waves arising from gravitational
instabilities in the gas are able to produce large scale
planetesimals, even if the effects of self-gravity in the gas is
relatively weak. Determination of the full extent of the region
where this mechanism can operate is beyond the scope of the present
simulations, since to probe inner radii down to about $10$AU, where
effective (radiative) cooling times are long but at the same time
Toomre's parameter is $Q\sim 1$, i.e., the effect of gas
self-gravity can still be appreciable \citep[e.g.,][]{Boley2006,
RiceArm2009, Cossins2010}, long simulation times are required, which
are not feasible for the simulations posed here. For such large
cooling times, numerical viscosity will begin to dominate the shear
stresses, requiring us to perform higher resolution studies.

In summary, the presented results tend to support and expand upon
those obtained in \citet{Rice2006} and Paper I, suggesting an
attractive scenario for the rapid creation of a reservoir of
planetesimals, along with several very massive objects with $\sim
0.01$ Earth masses. One of the main findings of this study has been
to demonstrate the possibility for this mechanism to form
planetesimals at a larger range of radii than previously thought
\citep[see e.g.,][]{ClarkeLodato2009}. This process potentially
solves a major problem in the standard planet formation scenario.
Rather than rapidly migrating into the central star, centimetre to
metre-sized particles become concentrated in self-gravitating spiral
structures. The densities achieved can then lead to planetesimal
formation via direct gravitational collapse of the particle
aggregates. In this scenario, kilometre-sized planetesimals form
very early, removing a major bottleneck in the planet formation
process. However, for a fuller understanding of the role of this
scenario in the planet formation, one should study the process of
subsequent growth and interaction of these planetesimals, which then
decouple from the gas and should be dominated by their own
gravitational attraction.

In this paper, we investigated the dynamics of gas and dust
particles in an idealized razor thin disc, so the limitations of
such a 2D model and its extension to the three-dimensional (3D) case
should be discussed. For the gaseous component, the description of
gravitational instabilities within the 2D shearing sheet model is
acceptable \citep[see e.g.,][]{Goldreich1978,Gammie2001}, since the
characteristic horizontal length scale of the instability and
induced structures (density waves) is larger than the vertical one
\citep{Goldreich1965,Romeo1992,Mamatsashvili2010,Shi2013}. As a
result, the gas motion associated with self-gravitating density
waves occurs primarily in disc plane. The situation is more
complicated with dust particles, since in the 2D case we cannot take
into account their motions perpendicular to the disc mid-plane, or
sedimentation, which depends on the particle size -- smaller
particles are well mixed with the gas, essentially do not sediment
and closely follow the gas, whereas particles with larger (from cm
to metre) size gradually settle towards the disc midplane on a
timescale of a few orbital times \citep{Goldreich1973}. This implies
that the back-reaction drag force from the particles on the gas,
which we calculated in terms of the ratio of the vertically
integrated surface densities of the particles and gas,
$\Sigma_p/\Sigma$, in equation (2), is strictly speaking valid if
particles are well-mixed with the gas. For larger particles, as they
settle into the midplane, the ratio of the the bulk density of
particles to the volume gas density, $\rho_p/\rho$, there is
expected to be larger than the ratio of the corresponding surface
densities, $\Sigma_p/\Sigma$, and since in 2D particles and the gas
have the same infinitely thin scale height, this causes the
back-reaction of the drag force from the particles on the gas to be
underestimated in the 2D case \citep{Youdin2005,Lyra2008b,Lyra2009}.
In the 3D case, the stronger back-reaction force on the gas from the
settled dust particles close to the midplane is known to lead to
streaming \citep{Youdin2005,Youdin2007} and Kelvin-Helmholtz
\citep{Sekiya1998,Youdin2002,Johansen2006} instabilities. The
streaming instability enhances particle clumping, thus aiding
collapse \citep{Johansen2007b}, while the Kelvin-Helmholtz
instability causes vertical stirring of the dust layer
\citep[e.g.,][]{Johansen2006}. To study in detail these 3D effects
related to particle dynamics in the presence of gas and particle
self-gravity and how they compete with the process of particle
trapping in density waves, one has to carry out more extensive
simulations in the 3D stratified shearing box. In the present local
analysis, following analogous 2D global simulations of particle-gas
dynamics by \citet{Lyra2008b,Lyra2009}, we have restricted ourselves
to expressing the back-reaction drag force by the particle and gas
column densities. Evidently, this is a simplification and meant as
an initial step towards understanding all the above complex
ingredients of particle dynamics in self-gravitating discs.
Nevertheless, such a 2D approach allows us to gain insight into the
characteristics of particle accumulation in overdense structures due
to self-gravity.

In regard to the above-mentioned, a question may arise as to whether
there is still a way to incorporate sedimentation of the particles
in the 2D model of gas-dust coupling. When the particles sediment,
the particle scale height is set by the balance between turbulent
stirring (diffusion) and vertical gravity
\citep[e.g.,][]{Johansen2005,Fromang2006}. Being controlled by the
drag force, the turbulent diffusion and therefore the equilibrium
scale height of solids $H_p$ depend on the particle radius (friction
time). As a consequence, larger particles settle into thinner layers
than smaller ones (obviously particle scale heights are different
from that of gas). Inside density waves, where, as mentioned above,
motion is horizontal, vertical turbulent motions are expected to be
weaker bringing the layer of solids closer to a 2D quasi-static
configuration, but a dependence on particle size is still expected.
Provided these scale heights of solids are known, one could, in
principle, find particle surface density as $\Sigma_p\approx
2\rho_pH_p$ and similarly for the gas $\Sigma\approx 2\rho H$ and in
this way relate the ratios of dust to gas volume and surface
densities, $\Sigma_p/\Sigma\approx (H_p/H)(\rho_p/\rho)$. However,
in the 2D case, $H_p$ remains largely uncertain, as it depends on
the vertical stirring properties of gravitoturbulence (and other
above-mentioned instabilities which will develop in 3D) and should
be self-consistently determined through 3D analysis.

\section*{Acknowledgments}
This work made use of the facilities of HECToR, the UK�s national
high-performance computing service, which is provided by UoE HPCx
Ltd at the University of Edinburgh, Cray Inc and NAG Ltd, and funded
by the Office of Science and Technology through EPSRC�s High End
Computing Programme. GRM acknowledges financial support from the
Rustaveli National Science Foundation (Georgia). WKMR acknowledges
support form STFC grant ST/J001422/1.

\end{document}